\documentclass[a4paper,12pt]{article}
\pdfoutput=1
\usepackage{amsmath}
\usepackage{amssymb}
\usepackage{mathrsfs}
\usepackage{bbm}
\usepackage{multirow}
\usepackage{graphicx,subfigure}
\usepackage[numbers,sort&compress]{natbib}

\usepackage{color}
\definecolor{brown}{cmyk}{0, 0.8, 1, 0.6}
\definecolor{orange}{rgb}{1,0.5,0}
\usepackage{ulem}

\numberwithin{equation}{section}

\newlength{\dinwidth}
\newlength{\dinmargin}
\makeatletter
\newcommand{\thickhline}{%
\noalign {\ifnum 0=`}\fi \hrule height 1pt
\futurelet \reserved@a \@xhline
}
\makeatother

\begin{document}

\title{\bf \Large 
Probing the Higgs sector of $Y=0$ Higgs Triplet Model at LHC}
\author{
M.~Chabab $^{a}$\footnote{mchabab@uca.ac.ma}, 
M.~C. Peyran\`ere $^{b}$\footnote{michel.capdequi-peyranere@umontpellier.fr},
L.~Rahili $^{a,c}$\footnote{rahililarbi@gmail.com}
}

\date{}
\maketitle
\vspace{-1.4em}
\begin{center}
\textit{$^a$
LPHEA, Faculty of Science Semlalia, Cadi Ayyad University, P.O.B. 2390 Marrakech, Morocco}\\[0.5em]
\textit{$^b$
LUPM, Montpellier University, F-34095 Montpellier, France}\\[0.5em]
\textit{$^c$
EPTHE, Faculty of Sciences, Ibn Zohr University, P.O.B. 8106 Agadir, Morocco}\\[0.5em]
\end{center}

\vspace{1em}

\begin{abstract}
{\noindent}
In this paper, we investigate the Higgs Triplet Model with hypercharge $Y_{\Delta}=0$ (HTM0),  an extension of the Standard model, 
caracterized by a more involved scalar spectrum consisting of two CP even Higgs $h^0, H^0$ and two charged Higgs bosons $H^\pm$. 
We first show that the parameter space of HTM0, usually delimited by combined constraints originating from unitarity and BFB as well as experimental limits from LEP and LHC, is severely reduced when the modified Veltman conditions at one loop are also imposed.  
Then, we perform an rigorous analysis of Higgs decays either when $h^0$ is the SM-like or when the heaviest neutral Higgs $H^0$ 
is identified to the observed $125$ GeV Higgs boson at LHC.  In these scenarios, we perform an extensive parameter scan, 
in the lower part of the scalar mass spectrum, with a particular focus on the Higgs to Higgs decay modes  $H^0 \to h^0h^0, H^\pm\,H^\mp$
leading predominantly to invisible Higgs decays. Finally, we also study the scenario where  $h^0, H^0$  are  mass degenerate. 
We thus find that consistency with LHC signal strengths favours a light charged Higgs with a mass about $176\sim178$ GeV.
 Our analysis shows that the diphoton Higgs decay mode and $H \to  Z \gamma$ are not always positively correlated as claimed in a previous study. Anti-correlation is rather seen in the scenario where $h$ is SM like, while correlation is sensitive to the sign of the potential parameter $\lambda$ when $H$ is identified to $125$ GeV observed Higgs.
\end{abstract}

\section{Introduction}
\label{intro}
Without a doubt, the neutral scalar boson discovered by  ATLAS \cite{atlas12} and CMS \cite{cms12} at the Large Hadron Collider (LHC)
corresponds to the Higgs boson. All data collected at $7$ and $8$ TeV  support the existence of Higgs signal with a mass around $125$ GeV
with  Standard Model (SM) like properties. Moreover, the deviation in $\gamma\gamma$ channel for the gluon and vector boson fusion productions, the Higgs production and decays into $ W W$* and $Z Z$* are all consistent with SM predictions, as can be seen from LHC run II measurements at 13 TeV \cite{atlas18,cms18}.
 
Similarly to our previous phenomenological analysis in the type II seesaw model \cite{aoki08,akeroyd10,aa11,chabab14,chabab16} we focus in this work on the Higgs Triplet Model with hypercharge $Y_{\Delta}=0$, hereafter referred to as HTM0. The main motivation of the HTM0 is related to the mysterious nature of dark matter (DM) and dark energy, which may signal new physics beyond the SM \cite{Chardonnet93,p.f.perez16,mariana2015}. Although a recent analysis of the HTM0 has been done in \cite{wang2014}, we revisit this model in light of new data at LHC run II, with the aim to  improve the previous analysis of the Higgs decays which suffered from some inconsistencies that produced inappropriate results for the correlation between Higgs to diphoton decay and Higgs to photon and a $Z$ boson.  
Furthermore,  our work will investigate the naturalness problem in HTM0. We will show how the new degrees of freedom in the HTM0 spectrum can soften the quadratic divergencies  and how the Veltman conditions are modified accordingly (VC) \cite{drozd,masina,kundu13,biswas14}. As a consequence, we will see that the parameter space of our model is severely constrained by the modified Veltman conditions.  

This paper is organised as following. In section 2, we briefly review the main features of HTM0, and present the full set of constraints on the parameters of the Higgs potential. Section 3 is devoted to the derivation of the modified VC's in HTM0. The Higgs sector is discussed in greater detail in section 4 where  either $h^0$ or $H^0$ are identified to the SM-like Higgs, and at last we focus on the scenario of their mass degeneracy where both Higgses  mimic the observed $\sim\,125$ GeV. A full set of  constraints were taken into account in the various analyses, including theoretical (BFB, unitarity) as well as the experimental ones, and scrutinised via HiggsBounds v4.2.1 \cite{bechtle2015v421} which we use to check agreement with all $2\sigma$ exclusion limits from LEP, Tevatron and LHC Higgs searches. Our conclusion is drawn in section 5, while some technical details are postponed into appendices.

\section{Review of the HTM0 model}
\label{sec:review}
\subsection{Lagrangian and Higgs masses}
\label{subsec:review}
The Higgs triplet model with hypercharge $Y_\Delta=0$ can be implemented in 
the Standard Model by adding a colourless scalar field $\Delta$ transforming as a 
triplet under the $SU(2)_L$ gauge group with hypercharge $Y_\Delta=0$. The most general gauge invariant and renormalisable
$SU(2)_L \times U(1)_Y$ Lagrangian of the scalar sector is given by,
\begin{eqnarray}
\mathcal{L} &=& (D_\mu{H})^\dagger(D^\mu{H})+Tr(D_\mu{\Delta})^\dagger(D^\mu{\Delta}) \nonumber\\
&-&V(H, \Delta)+\mathcal{L}_{\rm Yukawa}
\label{eq:DTHM}
\end{eqnarray}
\noindent
where the covariant derivatives are defined by, 
\begin{eqnarray}
D_\mu{H} &=& \partial_\mu{H}+igT^a{W}^a_\mu{H}+i\frac{g'}{2}B_\mu{H} \label{eq:covd1}\\
D_\mu{\Delta} &=&
  \partial_\mu{\Delta}+ig[T^a{W}^a_\mu,\Delta]. \label{eq:covd2}
\end{eqnarray}

\noindent
(${W}^a_\mu$, $g$), and ($B_\mu$, $g'$) are respectively the $SU(2)_L$ and $U(1)_Y$ gauge fields and couplings
and $T^a \equiv \sigma^a/2$, where $\sigma^a$ ($a=1, 2, 3$)  denote the Pauli matrices.
The potential $V(H, \Delta)$ can be expressed as \cite{p.f.perez16},
\begin{eqnarray}
\hspace{-0.5cm}V(H, \Delta)\hspace{-0.1cm}&=&\hspace{-0.1cm}- m_H^2{H^\dagger{H}} + \frac{\lambda}{4}(H^\dagger{H})^2 - 
M_\Delta^2Tr(\Delta^{\dagger}{\Delta})+ \mu\,H^{\dagger}\Delta H \nonumber\\
&&\hspace{-0.5cm}+\lambda_1(H^\dagger{H})Tr(\Delta^{\dagger}{\Delta})+\lambda_2(Tr\Delta^{\dagger}{\Delta})^2
+\lambda_3Tr(\Delta^{\dagger}{\Delta})^2 \nonumber\\
&&\hspace{-0.5cm}+\lambda_4{H^\dagger\Delta^{\dagger}\Delta H}
\label{eq:Vpot}
\end{eqnarray}
where $Tr$ is the trace over $2\times2$ matrices. Last, $\mathcal{L}_{\rm Yukawa}$ contains all the Yukawa sector
of the SM plus an extra Yukawa term that leads after spontaneous symmetry breaking to (Majorana) mass terms for
the neutrinos, without requiring right-handed neutrino states.

Defining the electric charge as usual, $Q= I_3 + \frac{Y}{2}$ where $I$ denotes the isospin, we write the two Higgs multiplets 
in components as:  
\begin{eqnarray}
\Delta &=\frac{1}{2}\left(
\begin{array}{cc}
\delta^0 & \sqrt{2}\delta^{+} \\
\sqrt{2}\delta^{-} & -\delta^0 \\
\end{array}
\right)~~~~{\rm and}~~~~H=\left(
                    \begin{array}{c}
                      \phi^+ \\
                      \phi^0 \\
                    \end{array}
                  \right)
\end{eqnarray}
with 
\begin{eqnarray}
\phi^0 &=\frac{1}{\sqrt{2}}(v_d + h_1 + i\,z_1)~~~~{\rm and}~~~~\delta^0=v_t + h_2
\label{neutre_components}
\end{eqnarray}
For later convenience, the vacuum expectation values $v_d$ and $v_t$ are supposed positive values.

Assuming that spontaneous electroweak symmetry breaking (EWSB) is taking place at some electrically 
neutral point in the field space, and denoting the corresponding VEVs by
\begin{eqnarray}
\langle \Delta \rangle &=\frac{1}{2}\left(
\begin{array}{cc}
v_t & 0 \\
0 & - v_t\\
\end{array}
\right)~~~~{\rm and}~~~~\langle H \rangle =\left(
                    \begin{array}{c}
                      0 \\
                       v_d/\sqrt{2}\\
                    \end{array}
                  \right)
\label{eq:VEVs}
\end{eqnarray}
\noindent
one finds, after minimisation of the potential Eq.(\ref{eq:Vpot}), the following necessary conditions :
\begin{eqnarray}
  M_\Delta^2 &=& \frac{\lambda_a}{2}v_d^2 - \frac{\mu{v_d^2}}{4 v_t}  + \lambda_b v_t^2 \label{eq:ewsb1}\\
  m_H^2 &=& \frac{\lambda}{4} v_d^2 - \frac{\mu v_t}{2} +\frac{\lambda_a}{2} v_t^2   \label{eq:ewsb2}
\end{eqnarray}
where $\lambda_a=\lambda_1+\frac{\lambda_4}{2}$ and $\lambda_b=\lambda_2+\frac{\lambda_3}{2}$.

The $7\times 7$ squared mass matrix,
\begin{equation}
 {\mathcal M}^2=\frac{1}{2} \frac{\partial^2 V}{\partial \eta_i^2} |_{\Delta = \langle \Delta \rangle , H= \langle H \rangle}
\end{equation}
\noindent
can be cast, thanks to Eqs.~(\ref{eq:ewsb1}, \ref{eq:ewsb2}), into a block diagonal form of three 
$2 \times 2$ matrices, denoted in the following by ${\mathcal{M}}_{\pm}^2, {\mathcal{M}}_{{\mathcal{CP}}_{even}}^2$, 
and one odd eigenstate corresponding to the neutral Goldstone boson $G^0$. The mass-matrix for singly charged field given by, 
$$
{\mathcal{M}}_{\pm}^2= \mu \left(
  \begin{array}{cc}
  v_t & v_d/2\\
  v_d/2 & v_d^2/4v_t\\
  \end{array}
\right)
$$
is diagonalised by a $2 \times 2$ rotation matrix
$\mathcal{R}_{\theta_{\pm}}$, where $\theta_{\pm}$ is a rotation angle. Among the two eigenvalues of ${\mathcal{M}}_{\pm}^2$, one is equal to zero 
indentifying the charged Goldstone boson $G^\pm$, while the other one corresponds to the mass of  singly charged Higgs bosons 
$H^\pm$ given by,  
\begin{eqnarray}
m_{H^\pm}^2=\frac{(v_d^2+4 v_t^2)}{4v_t} \mu
\label{eq:mHpm}
\end{eqnarray}
The mass-eigenstate $H^\pm$ and $G^\pm$ are rotated from the Lagrangian 
fields $\phi^{\pm}, \delta^{\pm}$ as follows :
\begin{eqnarray}
G^\pm &=& +\cos\theta_{\pm} \phi^{\pm}+ \sin\theta_{\pm} \delta^{\pm} \\
H^\pm &=& -\sin\theta_{\pm} \phi^{\pm}+ \cos\theta_{\pm}\delta^{\pm}
\end{eqnarray}
Diagonalization of ${\mathcal{M}}_{\pm}^2$  leads to the following relations involving the rotation angle $\theta_{\pm}$:
\begin{eqnarray}
\mu\frac{v_d^2}{4v_t} &=& \cos^2\theta_{\pm}M_{H^\pm}^2 \label{eq:sbp1}\\
\frac{\mu v_d}{2} &=& -\frac{\sin2\theta_{\pm}}{2}\,M_{H^\pm}^2 \label{eq:sbp2} \\
\mu v_t &=&  \sin^2\theta_{\pm}M_{H^\pm}^2 \label{eq:sbp3}
\end{eqnarray}
since the Goldstone boson $G^\pm$ is massless. These three equations have a unique solution for $\sin\theta_{\pm}$ and 
$\cos\theta_{\pm}$ up to a global sign ambiguity. Indeed,
Eq.~(\ref{eq:sbp1}) implies $\mu > 0$ in order to forbid tachyonic $H^\pm$ state, since our convention uses $v_t > 0$.
Hence, from Eq.~(\ref{eq:sbp2}), $\sin\theta_{\pm}$ and $\cos\theta_{\pm}$ should have different signs; one gets :
\begin{eqnarray}
\cos\theta_{\pm}&=& \epsilon \frac{v_d}{\sqrt{v_d^2+4v_t^2}}, \quad
\sin\theta_{\pm}= - \epsilon \frac{2 v_t}{\sqrt{v_d^2+4v_t^2}} \label{eq:scbprime}
\end{eqnarray}
\noindent
with a sign freedom $\epsilon=  \pm 1$, which leads to negative $\tan\theta_{\pm}$.

\noindent
As to the neutral scalar, its mass matrix reads:
\begin{eqnarray}
{\mathcal{M}}_{{\mathcal{CP}}_{even}}^2=\left(
                                                  \begin{array}{cc}
                                                   A & B \\
                                                   B & C  \end{array}
\right)
\label{cpeven:matrix}
\end{eqnarray}
where
\begin{eqnarray}
A = \frac{\lambda}{2}v_d^2,\,
B = \frac{v_d\,\big[-\mu + 2 \lambda_a\,v_t\big] }{2\sqrt{2}},\,
C = \frac{\mu v_d^2 + 8\lambda_b\,v_t^3}{8 v_t} 
\label{eq:ABC}
\end{eqnarray}
This symmetric matrix is also diagonalised by a $2 \times 2$ rotation matrix ${\mathcal{R}}_\alpha$,
\noindent
where $\alpha$ denotes the rotation angle in the ${\mathcal{CP}}_{even}$ sector.

After diagonalization of ${\mathcal{M}}_{{\mathcal{CP}}_{even}}^2$,  
one gets two massive even-parity physical states  $h^0$ and $H^0$ defined by,
\begin{eqnarray}
h^0 &=& +c_\alpha \, h_1 + s_\alpha \, h_2  \\
H^0 &=& - s_\alpha \, h_1 + c_\alpha \, h_2
\end{eqnarray}
\noindent
Their masses are given by the eigenvalues of  ${\mathcal{M}}^2_{{\mathcal{CP}}_{even}}$ :
\begin{eqnarray}
m_{h^0}^2&=&\frac{1}{2}[A+C-\sqrt{(A-C)^2+4B^2}] \label{eq:mh1}\\
m_{H^0}^2&=&\frac{1}{2}[A+C+\sqrt{(A-C)^2+4B^2}] \label{eq:mh2}
\end{eqnarray}
\noindent
so that $m_{H^0}>m_{h^0}$. Note that the lighter state $h^0$ is not necessarily the lightest of the Higgs sector. 
Furthermore, the only odd eigenstate leads to one massless Goldstone boson $G^0$ defined by $G^0 = z_1$. 

Once we know the above eigenmasses for the $\mathcal{CP}_{even}$, one can determine the rotation angle $\alpha$ which controls the 
field content of the physical states. One has : 
\begin{eqnarray}
C &=& s_\alpha^2m_{h^0}^2+c_\alpha^2m_{H^0}^2 
\label{eq:sa1}\\
B &=& \frac{\sin2\alpha}{2}(m_{h^0}^2-m_{H^0}^2) \label{eq:sa2}\\
A &=& c_\alpha^2m_{h^0}^2+s_\alpha^2m_{H^0}^2  \label{eq:sa3}
\end{eqnarray}
\noindent
Both Eq.~(\ref{eq:sa1}) and Eq.~(\ref{eq:sa3}) should be equivalent upon use of $s_\alpha^2 + c_\alpha^2=1$ and Eqs.~(\ref{eq:mh1}, \ref{eq:mh2}). Furthermore, $s_\alpha, c_\alpha$ also do not have definite signs, depending on the sign of $B$. The relative sign between $s_\alpha$ and $c_\alpha$ depends on the values of $\mu$ as can be seen from
Eqs.(\ref{eq:sa2}, \ref{eq:ABC}). While they will have the same sign and $\tan \alpha > 0$  for most of the allowed 
$\mu$ and $\lambda_1, \lambda_4$ ranges, there will be a small but interesting domain of small $\mu$ 
values and  $\tan \alpha < 0$. 

Finally, from Eqs.~(\ref{eq:sa1} - \ref{eq:sa3}), it is easy to express $\alpha$ in terms of $A, B$ and $C$ (Eqs .~(\ref{eq:ABC})) via :
\begin{eqnarray}
&& \sin2\alpha = \frac{2 B}{\sqrt{(A-C)^2+4B^2}} \quad{\rm and} \nonumber\\
&& \cos2\alpha = \frac{A-C}{\sqrt{(A-C)^2+4B^2}}
\label{eq:mixingCPeven}
\end{eqnarray}
%

\subsection{Constraints in the HTM0}
The full experimental validation of the HTM0 would require not only evidence for the neutral and charged Higgs states but also
the experimental values for the various field couplings in the gauge and matter sectors of the model. Crucial tests would then be 
driven by the predicted  correlations among these measurable quantities. For instance, the $\mu$ and $\lambda$'s parameters
can be easily  expressed in terms of the physical Higgs masses and the mixing angle $\alpha$ as well as the VEV's $v_d, v_t$, 
using equations (\ref{eq:mHpm}), (\ref{eq:sa1} - \ref{eq:sa3}). One finds
\begin{eqnarray}
\lambda_a &=& \frac{1}{v_tv_d}\big\{ \sqrt{2}s_{\alpha}c_{\alpha}(m_{h^0}^2-m_{H^0}^2) + \frac{2v_tv_d}{v_d^2+4v_t^2}m_{H^\pm}^2 \big\}\label{eq:lambda_a}\\
\lambda_b &=& \frac{1}{v_t^2} \big\{  s_{\alpha}^2m_{h^0}^2+c_{\alpha}^2m_{H^0}^2 - \frac{v_d^2}{2(v_d^2+4v_t^2)} m_{H^\pm}^2\big\}  \label{eq:lambda_b}\\
\lambda &=& \frac{2}{v_d^2}\{c_{\alpha}^2m_{h^0}^2+s_{\alpha}^2m_{H^0}^2\} \label{eq:lamd} \\
\mu     &=& \frac{4 v_t}{v_d^2+4v_t^2} m_{H^{\pm}}^2 \label{eq:mu}
\end{eqnarray}

\noindent
The remaining two Lagrangian parameters $m_H^2$ and $M_\Delta^2$ are then related to the physical
parameters through the EWSB conditions Eqs.~(\ref{eq:ewsb1}, \ref{eq:ewsb2}). 

In the Standard Model the custodial symmetry ensures that the $\rho$ parameter,
$\rho\equiv \frac{M_W^2}{M_Z^2\cos^2\theta_W}$, is $1$ at tree level. In the HTM0, it is clear that $\delta^0$ don't contribute to the Z boson mass, and one obtains the $Z$ and $W$ gauge boson masses readily from Eq.~(\ref{eq:VEVs}) and the kinetic terms in Eq.(\ref{eq:DTHM}) as
\begin{eqnarray}
M_Z^2&=& \frac{(g^2+{g'}^2) v_d^2}{4}   \label{eq:mZ}
     =\frac{g^2 v_d^2}{4c_w^2} \\
M_W^2&=&\frac{g^2(v_d^2+4v_t^2)}{4}  \label{eq:mW}
\end{eqnarray}
\noindent
Hence the modified form of the $\rho$ parameter is $\rho=\frac{v_d^2+4v_t^2}{v_d^2}$.

Since we are interested in the limit $v_t \ll v_d$, we rewrite
\begin{equation}
\rho = 1 + 4 \frac{v_t^2}{v_d^2} = 1 + \delta \rho
\end{equation}
\noindent
with $\delta \rho = 4 \frac{v_t^2}{v_d^2}  > 0$ and $\sqrt{v_d^2 + 4 v_t^2} =246$ GeV. \\
From a global fit to EWPO one obtains the $1\sigma$ result \cite{pdg16},
\begin{eqnarray}
\rho_0 = 1.0004^{+0.0003}_{-0.0004}
\end{eqnarray}
Consequently, in what follows, we adopt the bound
\begin{eqnarray}
\big(\frac{2v_t}{v_d}\big)^2 \lesssim 0.0006\,\quad {\rm or\,\,equivalently}\quad v_t \lesssim 3\,{\rm GeV}
\end{eqnarray}

\noindent
The positivity requirement in the singly charged sector, Eq.~(\ref{eq:mHpm}), along with our phase convention $v_t >0$, 
lead only to positive values of $\mu$. 
The tachyonless condition in the ${\mathcal{CP}}_{even}$ sector, Eqs.~(\ref{eq:mh1}, \ref{eq:mh2}), is somewhat more involved and reads :
\begin{eqnarray}
&\mu v_d^2 + 4 \lambda  v_d^2 v_t + 8\,\lambda_b\, v_t^3  > 0 & \label{eq:posh0H0-0}\\
&-2 \mu^2 v_t + \mu (\lambda v_d^2 + 8\,\lambda_a\,v_t^2)
     + 8 (\lambda\lambda_b -  \lambda_a^2) v_t^3  \label{eq:posh0H0-1}
    > 0 &
\end{eqnarray}
The first equation is actually always satisfied thanks to the positivity of $\mu$ and the boundedness from below conditions for the potential.
The second equation, quadratic in $\mu$, will lead to new constraints on $\mu$ in the form of an allowed range
\begin{equation}
\mu_{-} < \mu < \mu_{+} \label{eq:posh0H0}
\end{equation}
\noindent
The full expressions of $\mu_{\pm}$ are given by
\begin{equation}
\mu_{\pm} = \frac{8\,\lambda_a\,v_t^2 + \lambda v_d^2 \pm \sqrt{16\,\lambda\lambda_a v_d^2 v_t^2 + 64\,\lambda\lambda_b v_t^4 + \lambda^2 v_d^4 }}{4v_t}
\label{eq:posh0H0}
\end{equation}
Let us discuss their behaviours in the favoured regime $v_t \ll v_d$. In this case one finds a vanishingly small $\mu_{-}$ given by
\begin{equation}
\displaystyle\mu_{-} = 
(\lambda_a^2 - \lambda\lambda_b) \,\frac{8}{\lambda} \frac{v_t^3}{v_d^2}  
+ {\cal O}(v_t^4) \label{eq:muminusapprox}
\end{equation}
\noindent
and a large $\mu_{+}$ given by
\begin{equation} 
\displaystyle \mu_{+} =  \frac{\lambda}{2} \frac{v_d^2}{v_t} + 4\,\lambda_a\,v_t +{\cal O}(v_t^2).
\label{eq:muplusapprox}
\end{equation}
\noindent
Depending on the signs and magnitudes of the $\lambda$'s, lower bound $\mu>0$ (positivity of Eq.~(\ref{eq:mHpm})) or $\mu_{-}$ will overwhelm the others. Moreover, these no-tachyon bounds will have  eventually to be amended by taking into account the existing experimental exclusion limits. This is straightforward for the charged Higgs boson $H^\pm$, thus we define for later reference :
\begin{eqnarray}
\mu_{\rm min} &=& \frac{4 \, v_t }{v_d^2 + 4 v_t^2} \, ({m_{H^\pm}^2})_{\rm exp} \label{eq:mumin}
\end{eqnarray}
\noindent
where $({m_{H^\pm}})_{\rm exp}$ denotes the experimental lower exclusion limit for the charged Higgs boson mass. So $\mu$ must be larger than $\mu_{min}$
in order for the mass to satisfy this exclusion limit. 

Upon use of Eqs.~(\ref{eq:VEVs}, \ref{eq:ewsb1}, \ref{eq:ewsb2}) in Eq.(~\ref{eq:Vpot}) one readily finds that the value
of the potential at the electroweak minimum, $\langle V \rangle_{\rm EWSB}$, is given by:
\begin{equation}
\langle V \rangle_{\rm EWSB} = -\frac{1}{16}(\lambda v_d^4 + 
4\,\lambda_b\,v_t^4 + 2\,v_d^2 v_t (2\,\lambda_a\,v_t- \mu ))
\end{equation}

\noindent
Since the potential vanishes at the gauge invariant origin of the field space, $V_{H=0, \Delta =0}=0$, then
spontaneous electroweak symmetry breaking would be energetically disfavoured if  
$\langle V \rangle_{\rm EWSB} >0$. One can thus require as a first approximation  the naive bound on $\mu$ 
\begin{equation}
\displaystyle \mu < \mu_{\rm max} \equiv 
\frac{\lambda}{2} \frac{v_d^2}{v_t} + 2\,\lambda_a v_t +{\cal O}(v_t^2) 
\label{eq:mumax}
\end{equation}

The phenomenological analysis in section $4$ is performed in the parameter space scanned by the potential parameters obeying the usual theoretical constraints, namely perturbative unitarity and boundedness form below (BFB). No need to mention that only the scan points that pass all these constraints are considered in our plots. \\

{\sl \underline{BFB}:}  \\
\noindent
To derive the BFB constraints, we usually consider that the scalar potential, at large field values, is generically dominated by its quartic part :
\begin{eqnarray}
V^{(4)}(H, \Delta) &=& \lambda(H^\dagger{H})^2/4
+\lambda_1(H^\dagger{H})Tr(\Delta^{\dagger}{\Delta})\nonumber\\
&+&\lambda_2(Tr\Delta^{\dagger}{\Delta})^2+\lambda_3Tr(\Delta^{\dagger}{\Delta})^2
+\lambda_4{H^\dagger\Delta^\dagger\Delta H}\nonumber\\
\label{eq:Vquartic}
\end{eqnarray}
In this context, it is common to pick up specific field directions or to put some of the couplings to zero. To proceed to the most general case, we adopt the same parameterisation as in \cite{aa11}, where in our model the $\xi$ and $\zeta$ parameters are found to be, 
\begin{equation}
\xi = \frac{1}{2}  \quad {\rm and} \quad  \zeta = \frac{1}{2}
\label{eq:conditions}
\end{equation}
The boundedness from below is then equivalent to requiring $V^{(4)} > 0$ {\sl for all} directions. As a result, the following set of conditions is derived:  
\begin{eqnarray}
&& \lambda \geq 0 \;\;{\rm \&}\;\; \lambda_b \geq 0  \label{eq:bound1} \\
&& {\rm \&} \;\;\lambda_a + \sqrt{\lambda\lambda_b} \geq 0 \label{eq:bound2}  
\end{eqnarray}

\noindent

{\sl \underline{Unitarity} \cite{nkhan2016}:}\\
\noindent 
As for unitarity constraints, they are given by,
\begin{eqnarray}
&&|\lambda_a| \leq \kappa\pi  \label{eq:unit1} \\
&&|\lambda| \leq 2\kappa\pi \label{eq:unit2} \\
&&|\lambda_b| \leq \frac{\kappa}{2} \pi \label{eq:unit3} \\
&&| 3 \lambda + 10 \lambda_b \pm \sqrt{(3 \lambda - 10 \lambda_b)^2
+ 48 \lambda_a^2} \;| \leq  4 \kappa \pi \label{eq:unit4} 
\end{eqnarray}

The details of their derivation are presented in appendix A.
Note that the parameter $\kappa$ is fixed to the value
values $\kappa=8$, since the unitarity formula $| Re(a_0)| \le \frac{1}{2}$ has been used.\\

\noindent
At this stage, by working out analytically these two sets of BFB and unitarity constraints, we can reduce them to a more compact system where the allowed ranges for the $\lambda$'s are easily identified. One can obtain a necessary domain 
for $\lambda, \lambda_b$ that does not depend on $\lambda_a$, by considering
simultaneously Eqs.~(\ref{eq:unit2} - \ref{eq:unit4}) together with Eq.~(\ref{eq:bound1}), 
\begin{eqnarray}
&& 0 \leq \lambda \leq \frac{2\kappa}{3} \pi  \label{eq:unit4new} \\
&& 0 \leq \lambda_b \leq \frac{\kappa}{5} \pi \label{eq:bound} \\
&& |\lambda_a| \leq \sqrt{\frac{5}{2} (\lambda - \frac{2}{3}\kappa\pi) (\lambda_b 
- \frac{\kappa}{5} \pi)} \label{eq:unit8new}
\end{eqnarray}

\noindent
We stress here that the above
constraints define the largest possible domain for $\lambda, \lambda_b$ for any set of allowed values of $\lambda_a$.   -Note also that, by using Eqs.~(\ref{eq:unit4new}-\ref{eq:bound}), one can rewrite Eq.~(\ref{eq:unit4}) under the simple form, given by Eq.~(\ref{eq:unit8new}), where the dependence on $\lambda_a$ has been explicitly separated from that on $\lambda, \lambda_b$.

The reduced couplings $g_{\mathcal{H} ff}$
and $g_{\mathcal{H} VV}$ of the Higgs bosons to fermions and $W$ bosons are
given in Tab.\ref{table_couplings}, while the trilinear couplings to charged Higgs bosons
can be extracted from the Lagrangian as  
${\mathcal{L}}= g_{{\mathcal{H}} H^{\pm}H^{\mp}}\mathcal{H}H^+H^-+g_{Z H^{\pm}H^{\mp}}Z(\partial_\mu H^+)H^-+\dots$. We will use the reduced HTM0 trilinear coupling of $\mathcal{H}$ and $Z$ to $H^\pm$ given by:
\begin{eqnarray}
&& g_{Z H^{+}H^{-}} = \frac{e}{2\,s_w\,c_w}(1-2\,c_w^2)\,s^2_{\theta_{\pm}} \nonumber\\
&& \tilde g_{\mathcal{H} H^{+}H^{-}}= -\frac{s_w}{e}\frac{m_W}{m_{H^+}^2}g_{\mathcal{H} H^+H^-}
\label{trilinear_coup1}
\end{eqnarray}

where $e$ is the electron charge,  $s_W$ the sinus of the weak mixing angle, and $m_W$ the mass of the gauge boson $W$.

\begin{table}[!ht]
\begin{center}
\begin{tabular}{|c|c|c|c|} 
\hline\hline
 $\cal{H}$  & $g_{\mathcal{H} ff}$ & $g_{\mathcal{H} WW}$ & $g_{\mathcal{H} ZZ}$ \\
\hline\hline
$h^0$  &  \ $ \; c_\alpha/c_{{\theta_{\pm}}} \; $ \ & \ $ \; c_{{\theta_{\pm}}} c_\alpha - 2 s_{{\theta_{\pm}}} s_\alpha \; $ \ &  \ $ \; c_{{\theta_{\pm}}} c_\alpha \; $ \ \\
$H^0$  &  \ $ \; - s_\alpha/ c_{{\theta_{\pm}}} \; $ \ & \ $ \; -c_{{\theta_{\pm}}} s_\alpha - 2 s_{{\theta_{\pm}}} c_\alpha \; $ \ &  \ $ \; -c_{{\theta_{\pm}}} s_\alpha \; $ \\ 
\hline\hline
\end{tabular}
\end{center}
\caption{ The CP-even neutral Higgs couplings to 
fermions and gauge bosons in the HTM0 {\sl relative} to the SM Higgs couplings.
$\alpha$ and $\theta_{\pm}$ are the mixing angles respectively in the CP-even and
charged Higgs sectors.}
\label{table_couplings}
\end{table}

The trilinear coupling $g_{h^0 H^+H^-}$ for the light CP-even Higgs boson is given by :
\begin{eqnarray}
g_{h^0 H^+ H^-} &=& -\frac{1}{2} \bigg\{c_\alpha\,(-2\,c_{\theta_{\pm}}\,s_{\theta_{\pm}}\,\mu + 
2\,\lambda_a\,c_{\theta_{\pm}}^2\,v_d + \lambda\,s_{\theta_{\pm}}^2\,v_d) \nonumber\\
&&\hspace{-0.6cm} + s_\alpha\,(4\,\lambda_b\,c_{\theta_{\pm}}^2\,v_t + s_{\theta_{\pm}}^2 (\mu + 2 \lambda_a v_t))\bigg\}
\label{trilinear_couph0}
\end{eqnarray}
The couplings for the heavy Higgs boson are obtained from the previous ones by simple substitutions 
$g_{H^0H^+H^-} = g_{h^0H^+H^-}  [c_\alpha \rightarrow -s_\alpha, s_\alpha \rightarrow c_\alpha]$.

\section{Veltman conditions}
To derive the Veltman conditions (VC), one just has to collect the quadratic divergencies \cite{Veltman81}. There are 
various ways to do that, and to be on a safer side, we use the dimensional regularisation because 
this procedure ensures  gauge as well as Lorentz invariances. To work out these quadratic divergencies, we follow exactly the procedure
of calculations used in our previous work on the Higgs Triplet Model with hypercharge $Y=2$ \cite{chabab16}. Moreover, it is worth 
to note that the main difference with \cite{chabab16} is the absence of the CP odd neutral Higgs $A^0$ and the doubly charged Higgs $H^{\pm\pm}$, 
from HTM$0$ spectrum.
Also we have calculated the quadratic divergencies of the CP-neutral Higgs 
$H^0$ and $h^0$ tadpoles in a general linear $R_{\xi}$ gauge respectively, leading to results which are independent of the  
$\xi$ parameters but depending on the model mixing angles. As noted in \cite{chabab16}, it is more convenient to combine these two results
to get the tadpoles quadratic divergencies of the real neutral components of the doublet ($h_1$) and triplet ($h_2$) which are
free of any mixing angles. After their VEV shifts, one finds, for the doublet: 
$$ T_d= v_d\,\Big(-2 Tr(I_n) \Sigma_f \frac{m_f^2}{v_d^2} +3 (\lambda + \lambda_a)
              +2  \frac{m_W^2}{v_{sm}^2} (\frac{1}{c_w^2} +2) \Big) $$
where $Tr(I_n)$ is the trace of the n-dimensional identity Dirac matrix, that is $2^{\frac{n}{2}}=2$ in our case. .

\noindent
For the triplet, one gets :
$$ T_t= v_t\,\Big(8 \frac{m_W^2}{v_{sm}^2} +2 \lambda_a + 5 \lambda_b\Big) $$

In the above expressions, we used the following simplified notations: $c_w = \cos \theta_{W}$ and $v_{sm} = \sqrt{v_d^2 + v_t^2}$.

Notice that the quadratic divergencies of the Standard Model are easily recovered in $T_d$ when the $\lambda_1$ and $\lambda_4$ couplings vanish, implying $\lambda_a =0$. 

Now to proceed with the implementation of the two VC's in the parameter space and the subsequent scans, we usually assume 
that the deviations $ \delta T_t$  and $\delta T_d $ should not exceed the Higgs mass scale. In our analysis, we will allow them 
to vary within the reduced conservative range from $0.1$ to $10$ GeV.

\begin{figure*}[t!]
\centering
\resizebox{0.4\textwidth}{!}{
\includegraphics{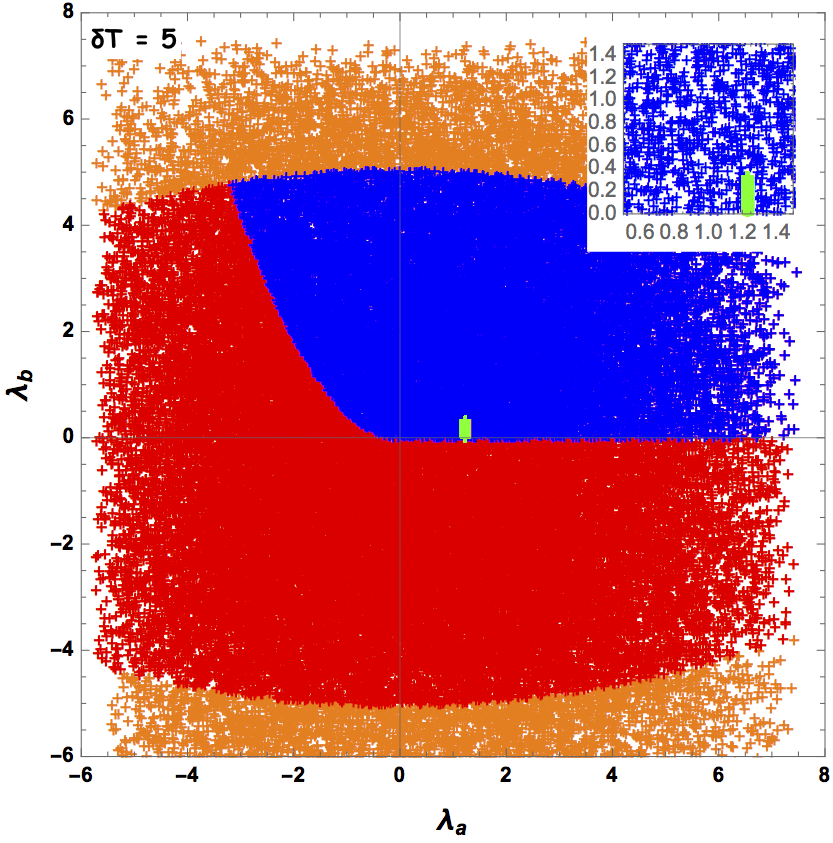}}
\resizebox{0.4\textwidth}{!}{
\includegraphics{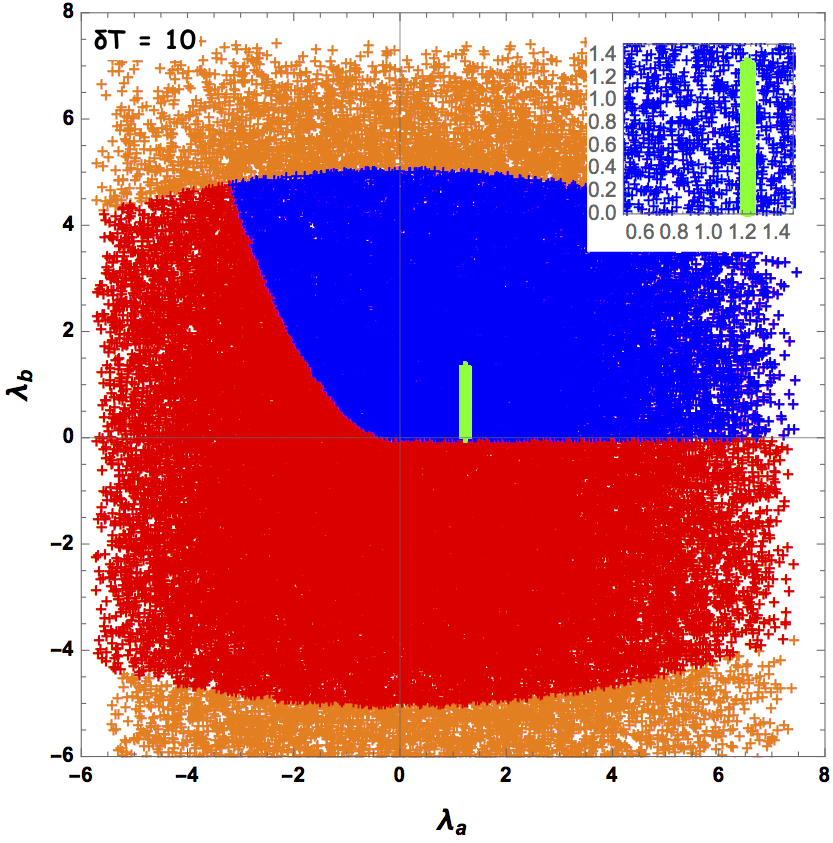}}
\caption{The allowed region in ($\lambda_{a},\lambda_{b}$) for two values of $\delta T = 5\,,10$. Color codes are as follows, 
\textcolor{orange}{Orange} : Excluded by Unitarity constraints. \textcolor{red}{Red} : Excluded by Unitarity \& BFB constraints.  
\textcolor{blue}{Blue} : Excluded by Unitarity \& BFB \& VC constraints.
The \textcolor{green}{Green} area represents the ALLOWED region of the parameter space obeying to all theoretical constraints. Our inputs are:  $\lambda = 0.52$, 
$-5 \le \lambda_1, \lambda_2, \lambda_3, \lambda_4, \lambda_5 \le 5$, $v_t=1$ GeV and $2\le \mu \le 5$ (GeV).}
\label{fg:fig1} 
\end{figure*}

\noindent In addition to the  theoretical constraints shown in Eqs.~\ref{eq:bound2}-\ref{eq:unit4}, namely the unitarity, BFB and
$R_{\gamma \gamma}$ from LHC measurements,  if the supplementary VC constraints are imposed as well,  we see that the allowed region
of the parameter space dramatically reduces and its extent depends on the value given to the deviation $ \delta T$. This salient
feature is illustrated in Fig.\ref{fg:fig1}, which exhibits the allowed domain in the $(\lambda_a,\lambda_b)$ plan. Our analysis shows
that naturalness constraint is stronger than the other theoretical conditions and that deviations $\delta T$ 
should be larger than $3$ GeV
in order to keep a viable model. Moreover, taken those constraints together, one might see that $\lambda_a$ will be restricted around 
$\sim\,1.2$, irrespectively of the value given to the {\it vev} $v_t$.
Indeed the same trends described above are reproduced when varying the triplet {\it vev}, though the $\lambda_a$ is somehow  freezer out.

Given the above discussed feature, in the next section, our phenomenological analysis will be performed within larger regions of parameter space that omit the VC constraints.

\section{Results and Discussions}
Since HTM$0$ spectrum contains two CP even Higgs boson $h^0$ and $H^0$, either $h^0$ or $H^0$ can be identified as the observed SM-like boson with mass $\approx 125$GeV. Therefore, we are facing two choices: $M_h^0\approx 125$ and  $M_h^0 \leq M_H^0$, or $M_H^0\approx 125$ and  $M_h^0 \leq M_H^0$.  For the former scenario, the mixing angle limit must verify $\cos\alpha \ge 0.96$, whereas when $H^0$ mimics the observed boson $\cos\alpha$ tends to a tiny value, so to keep consistency with the experimental data, we imposed $\sin\alpha\ge 0.96$. The third scenario considered in this paper is when both Higgs bosons are mass degenerate, $M_h^0 \approx M_H^0$. 

For evaluating the branching ratios we have taken into account the leading perturbative QCD corrections to the two CP-even Higgs decays into hadronic two-body final states. For the Higgs to diphoton and photon+Z gauge boson signal strengths, $R_{\gamma\gamma}$ and $R_{Z\gamma}$, we use the definition adopted in \cite{arhrib2012}, 
\begin{equation}
R_{\gamma\gamma(Z\gamma)}(\phi) = \frac{\Gamma_{\phi\to\,gg}^{HTM}\times\,BR_{\phi\to\gamma\gamma(Z\gamma)}^{HTM}}{\Gamma_{H\to\,gg}^{SM}\times\,BR_{H\to\gamma\gamma(Z\gamma)}^{SM}} 
\label{eq:RgagaRgaZ-def}
\end{equation}
The relevant ratios for the other channels $b\bar{b}$, $\tau^+\tau^-$, $W^+W^-$ and $ZZ$ are defined in a similar way. For the constraints and bounds from their corresponding signal strength measurements, we require agreement with the ATLAS and CMS at least at $1 \sigma$ ( see Appendix C for compilation of these signal strengths).  Our analyse shows that their ratios remain compatible with respect to its SM values since their $\mathcal{H}$ couplings are almost $\approx 1$. \\

Also, It should also be noted that for the CP-even Higgs
decays to final states with $b$ quarks, the QCD corrections up to three-loops have been included in their partial decay widths \cite{Chetyrkin:1995pd},
\begin{eqnarray}
\Gamma_{ {\mathcal H} \to q\bar{q}} &=&  \frac{3 G_F m_{\mathcal H}}{4 \sqrt{2} \pi} \bar{m}_{q}^2(m_{\mathcal H}) C_{qq}^{\mathcal H}\;{\rm\Delta}_{\rm QCD}
\label{eq:Gammabbar}
\end{eqnarray} 
where 
\begin{eqnarray}
{\rm\Delta}_{\rm QCD}&=&\Big( 1 + 5.67 \frac{\alpha_s (m_{\mathcal H})}{\pi} + (35.94 - 1.36
N_F) \frac{\alpha_s^2 (m_{\mathcal H})}{\pi^2}\nonumber\\
&+&(164.14 - 25.77 N_F + 0.259 N_F^2) \frac{\alpha_s^3(m_{\mathcal H})}{\pi^3} \Big)
\end{eqnarray} 

\begin{figure*}[t!]
\centering
\resizebox{0.38\textwidth}{!}{
\includegraphics{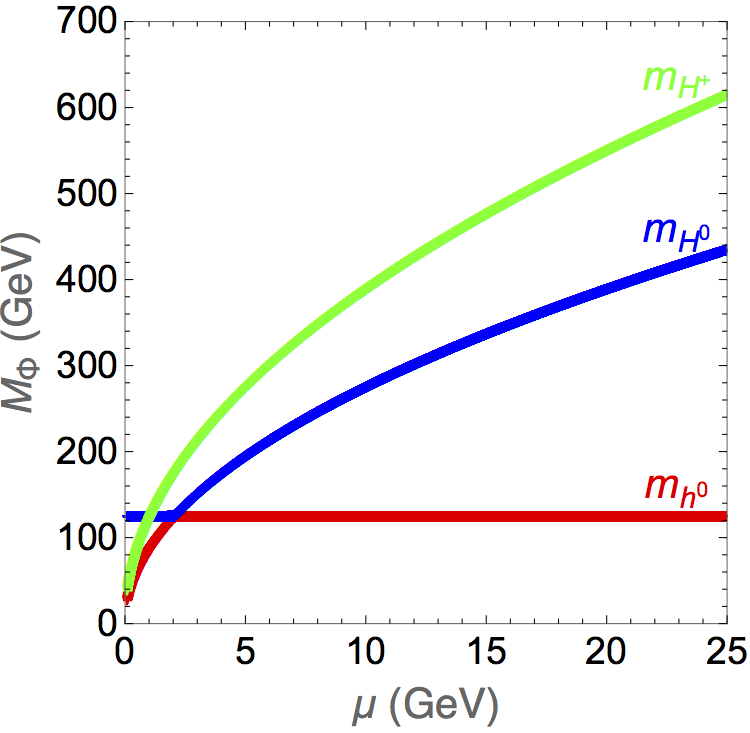}}
\resizebox{0.38\textwidth}{!}{
\includegraphics{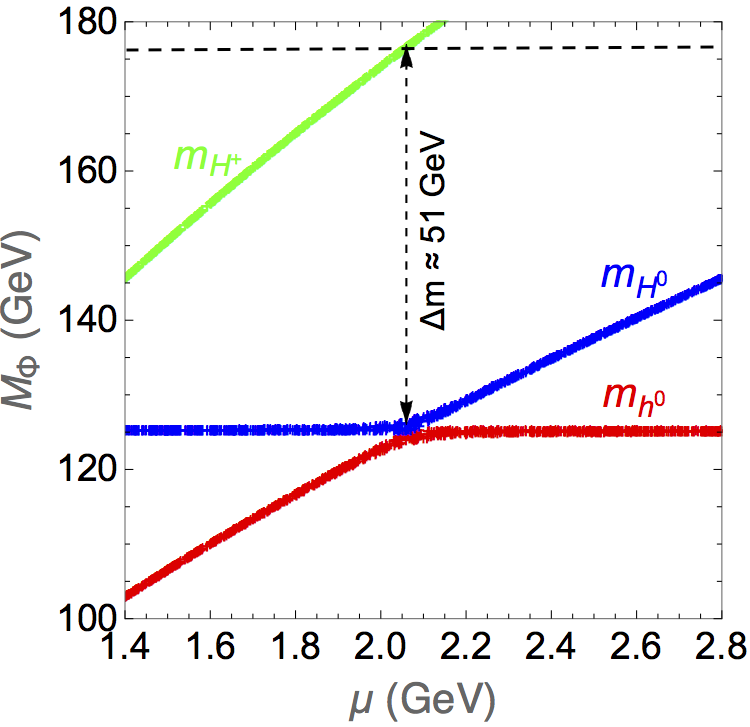}}
\caption{Higgs bosons masses as a function of $\mu$ parameter in the HTM$0$. We take as inputs $\lambda = 0.52$, 
$-1 \le \lambda_a \le 1$, $\lambda_b = 1$, $v_t = 1$ GeV and $0.1 \le \mu \le 25$ (GeV).}
\label{fg:fig2}
\end{figure*}

\begin{figure*}[t!]
\centering
\resizebox{0.32\textwidth}{!}{
\includegraphics{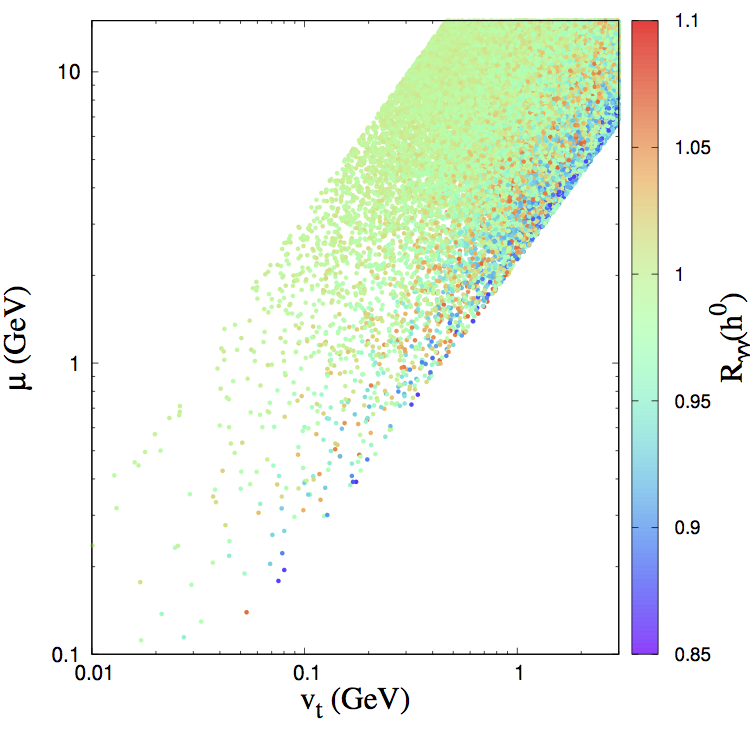}}
\resizebox{0.32\textwidth}{!}{
\includegraphics{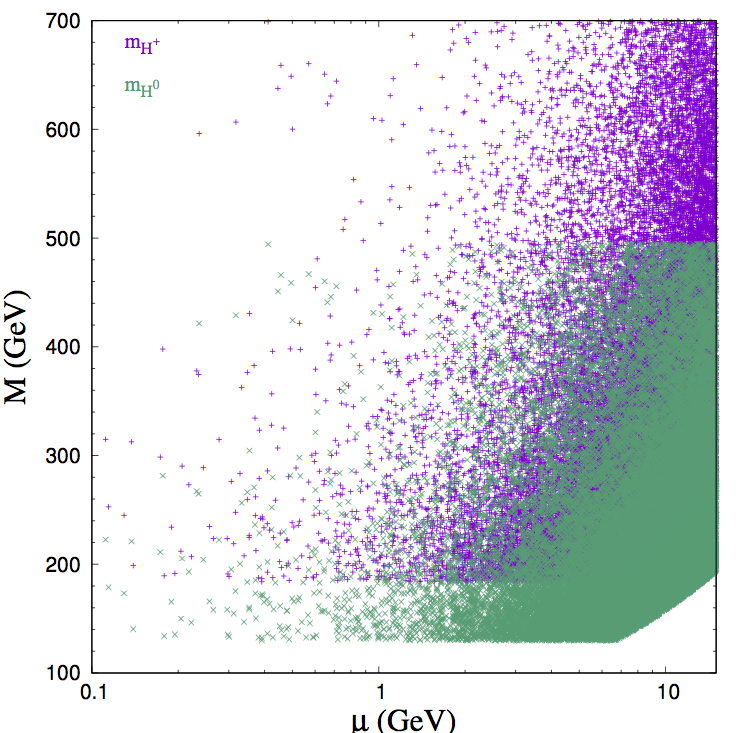}}
\resizebox{0.32\textwidth}{!}{
\includegraphics{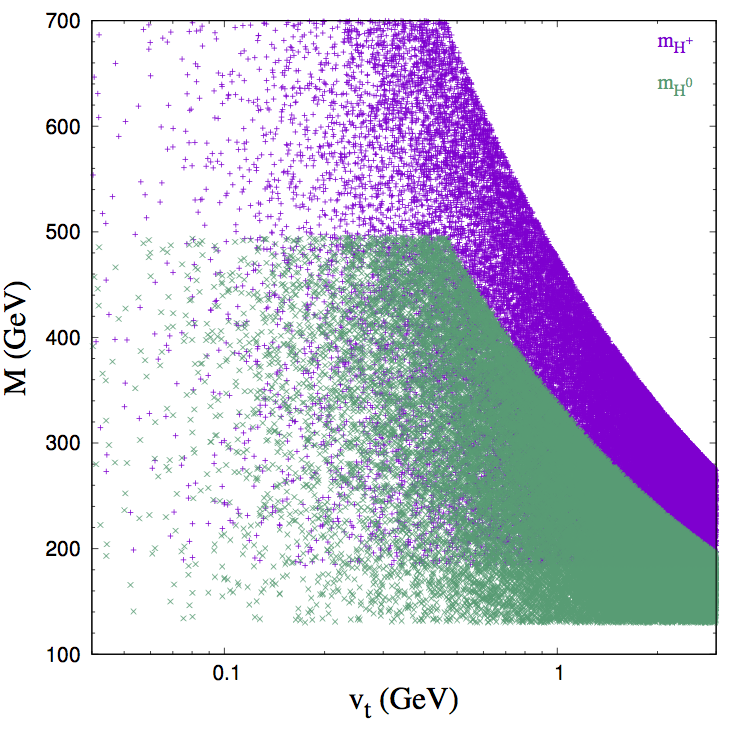}}
\caption{$R_{\gamma\gamma}(h^0)$ variation in the ($\mu,v_t$) plane (left), $H^0$ and $H^\pm$ Higgs bosons masses as a function of $\mu$ (middle) and of $v_t$ (right). Inputs are: $\lambda \approx 0.52$ ($m_{h^0} \approx 125$ GeV), $| \lambda_{a} | \le 1.5$, $| \lambda_{b} | \le 1$, $10^{-2} \le \mu \le 25$ (GeV)  and $10^{-2} \le v_t \le 3$ (GeV)}
\label{fg:fig3}
\end{figure*}

For each benchmark scenario, we investigate the allowed parameters space by the $1\sigma$ limit of the current Higgs data after run-II in the $gg \to \mathcal{H} \to\, \gamma\gamma$ channel, reported by ATLAS $\mu_{\gamma\gamma} = 0.85^{+0.22}_{-0.20}$ \cite{atlas_2016aug,atlas_2017myr,atlas_2017ovn} and CMS $\mu_{\gamma\gamma} = 1.11 ^{+0.19}_{-0.18}$ \cite{cms_2017jkd}, which are consistent with the Standard Model expectation either for ATLAS or for CMS at $1 \sigma$. It is worth noting that the errors reported here are smaller than those reported at $7\oplus8$ TeV.

\subsection{$h^0$ SM-like \label{sec:lightScalar}}
\noindent
Fig.~\ref{fg:fig3} displays the allowed region in the $(v_t,\mu)$, $(\mu,m_{H^0})$ and $(\mu,m_{H^\pm})$ planes, where $h^0$ is chosen to be SM-like. It is interesting to note that significant amount of parameter space is allowed once we impose either theoretical or experimental constraints, even for small nonzero value of $v_t$ and $\mu$.

\begin{figure*}[t!]
\centering
\resizebox{0.38\textwidth}{!}{
\includegraphics{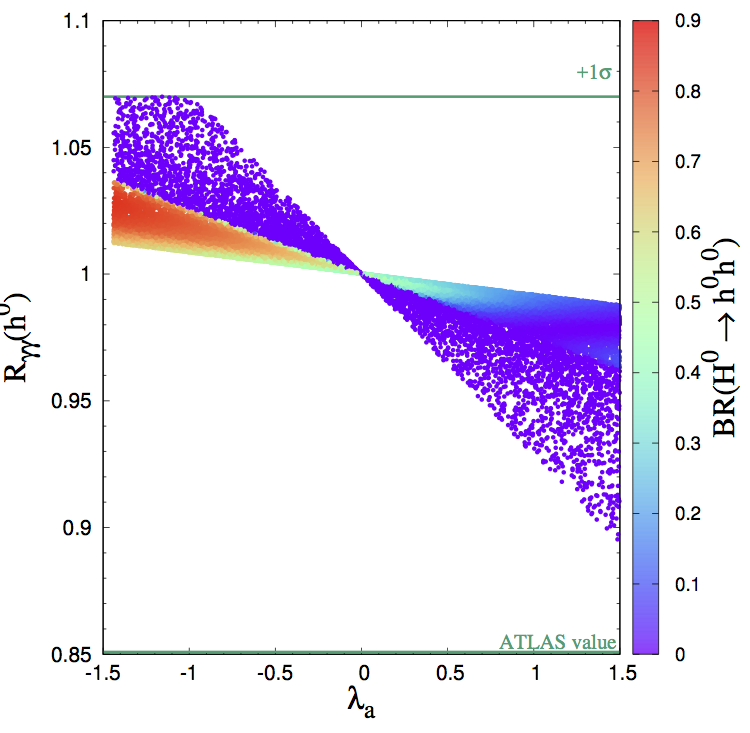}}
\resizebox{0.38\textwidth}{!}{
\includegraphics{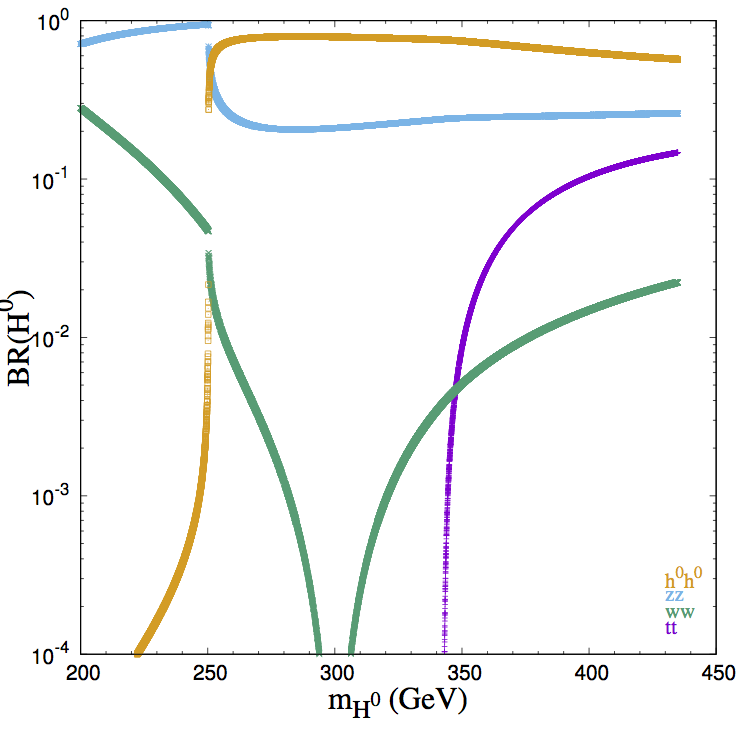}}
\caption{${\rm BR}(H^0 \to h^0h^0)$ variation in the ($R_{\gamma\gamma}(h^0)$, $\lambda_a$) plane taking into account ATLAS result at $1\sigma$, with the following inputs : $\lambda \approx 0.52$, $| \lambda_{a} | \le 1.5$, $\lambda_{b} =1$, $5 \le \mu \le 25$ (GeV) and $v_t = 1$ GeV (left). The ${\rm BR}(H^0)$ as a function of $m_{H^0}$  for a benchmark point where $\lambda_a=-1$ (right) }
\label{fig:fig4}
\end{figure*}

In order to establish in this case the branching ratios of the heaviest CP even neutral Higgs boson, we present in Fig.~\ref{fig:fig4} (right) the decay branching fractions of the heavier Higgs boson $H^0$ in the HTM0, for a benchmark point where $\lambda_a=-1$. We see that for $200\,{\rm GeV} < m_{H^0} < 250\,{\rm GeV}$, the dominant decay channels are the $H^0 \to Z Z$ and $W^+ W^-$ decay modes, whereas $h^0h^0$ is off-shell and consequently its corresponding ratio gets a tiny values of order of $1\%$, regardless of what $\lambda_a$ can be. Once  $h^0h^0$ threshold takes place, this channel becomes predominant for negative $\lambda_a$, with the ratio $R_{\gamma\gamma}(h^0)$  almost equal to its standard value, and $250 \le m_{H^0}< 450$  (GeV). This feature persists even when $t\bar{t}$ threshold is reached at $m_{H^0} > 350$ GeV.

\begin{figure*}[t!]
\centering
\resizebox{0.37\textwidth}{!}{
\includegraphics{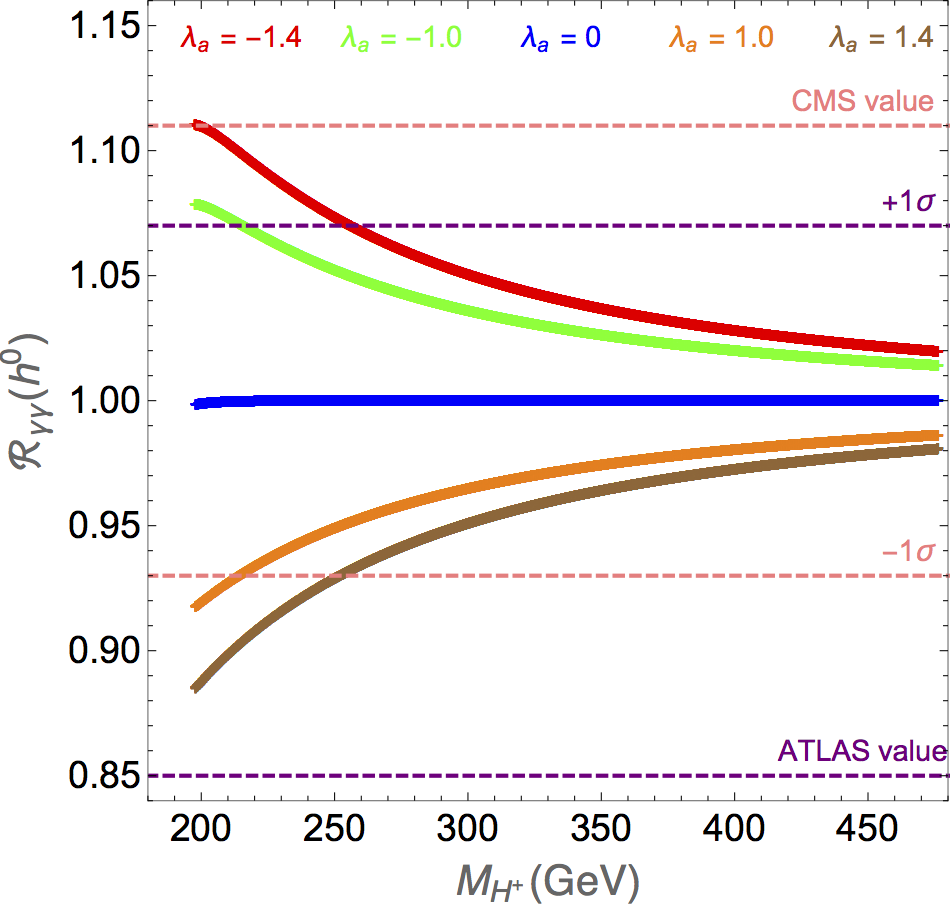}}
\resizebox{0.38\textwidth}{!}{
\includegraphics{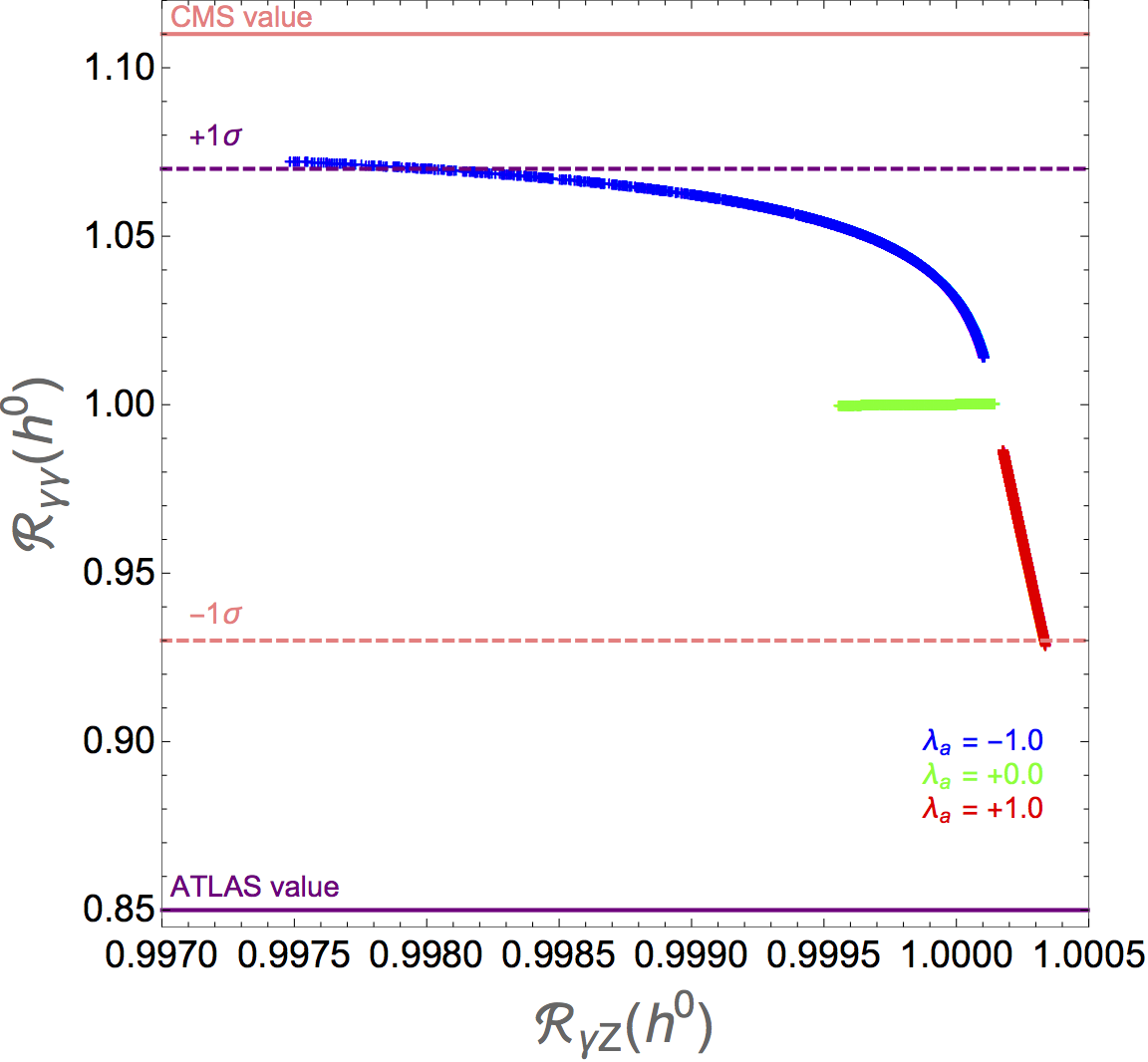}}
\caption{$R_{\gamma\gamma}(h^0)$ as a function of $m_{H^\pm}$ for various values of $\lambda_a$ (left). Correlation between $R_{\gamma\gamma}(h^0)$ and $R_{\gamma\,Z}(h^0)$ for various of $\lambda_a$ (right). We take as inputs : $\lambda \approx 0.52$, $2.5 \le \mu \le 15$ (GeV)  ($m_{h^0}\approx 125$ GeV), $\lambda_{b} =1$ and $v_t = 1$ GeV.}
\label{fig:fig5}
\end{figure*}

According to Eq.~\ref{eq:RgagaRgaZ-def}, we display the deficit of $R_{\gamma\gamma}(h^0)$ in the left panel of Fig.~\ref{fig:fig5} as a function of $H^\pm$ mass for various values of $\lambda_a$ and with $m_{H^0} \ge140$ GeV. As it can be seen, a mass about $255$ GeV and above is allowed for $H^\pm$ within $+1\sigma$ of ATLAS value for $\lambda_a=-1.4$. Once $\lambda_a$ increases, this lower bound decreases consistently to reach its lowest value around $\sim\,197$ GeV, given $\lambda_a >-0.5$. This situation is exactly the opposite for CMS, where only the range $200\le m_{H^\pm} \le250$ (GeV) is excluded for $\lambda_a=1.4$. Besides, $R_{\gamma\gamma}(h^0)$ tends towards its standard value for $\lambda_a\ne0$, and to 1 for large $m_{H^\pm}$ whatever the variation of $\lambda_a$.

In this scenario, the anti-correlation between $R_{\gamma\gamma}(h^0)$ and $R_{\gamma\,Z}(h^0)$ is displayed in the left panel of Fig.~\ref{fig:fig5}, taking into account the experimental data at $1\sigma$. At first sight, the $R_{\gamma\,Z}(h^0)$ deviation is almost nul relatively to its standard value, and  contrary to what has been claimed in \cite{wang2014}, $R_{\gamma\gamma}(h^0)$ and $R_{\gamma\,Z}(h^0)$ are always anti-correlated, independently of $\lambda_a$ sign.

\subsection{$H^0$ SM-like invisible decays \label{sec:invisible}}

This section investigates the possible existence of a scalar state $h^0$ lighter than $H^0$, with $M_H^0\approx 125$. Such a scenario has attracted attention within a plethora of theoretical frameworks dealing with new physics beyond standard model , particularly those considering enlargement of the Higgs sector of the SM via doublet or triplet fields \cite{arhrib2014,invisible}. However, to our knowledge,  it has not been addressed yet in the HTM0.

\begin{figure*}[t!]
\centering
\resizebox{0.38\textwidth}{!}{
\includegraphics{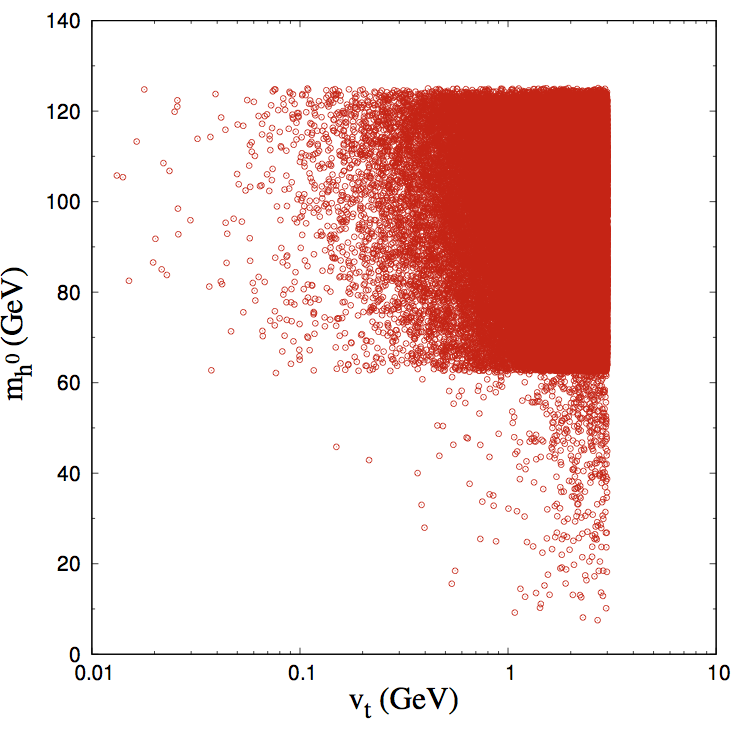}}
\resizebox{0.38\textwidth}{!}{
\includegraphics{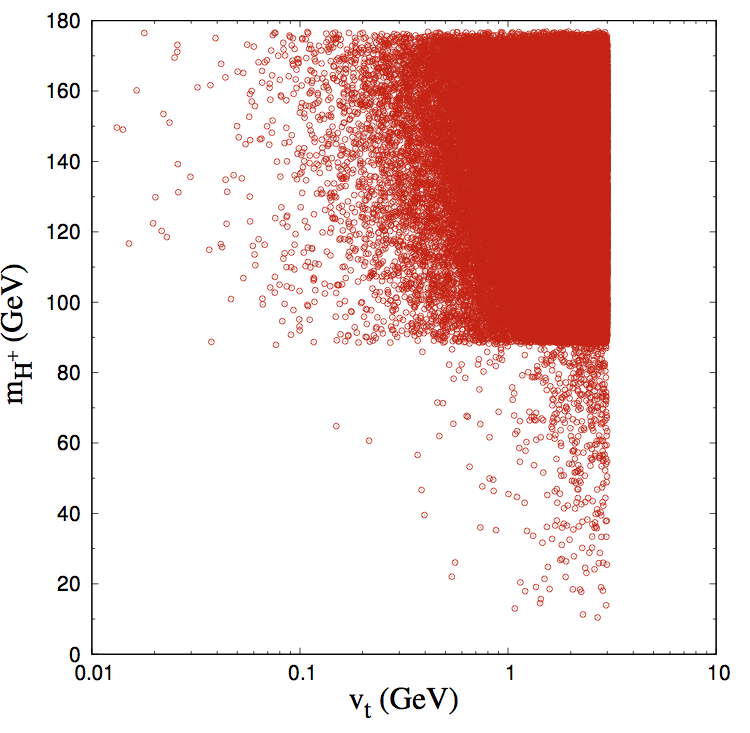}}
\resizebox{0.38\textwidth}{!}{
\includegraphics{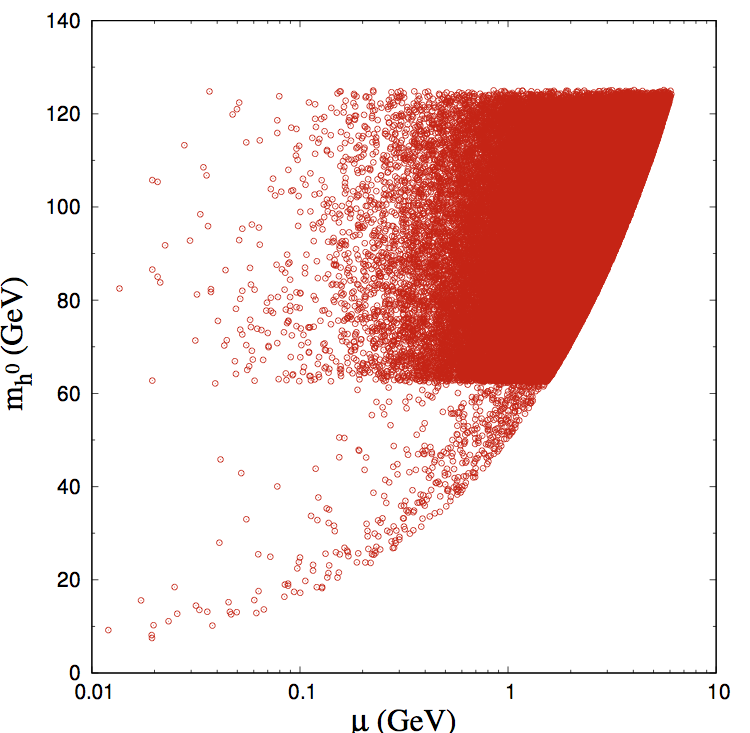}}
\resizebox{0.38\textwidth}{!}{
\includegraphics{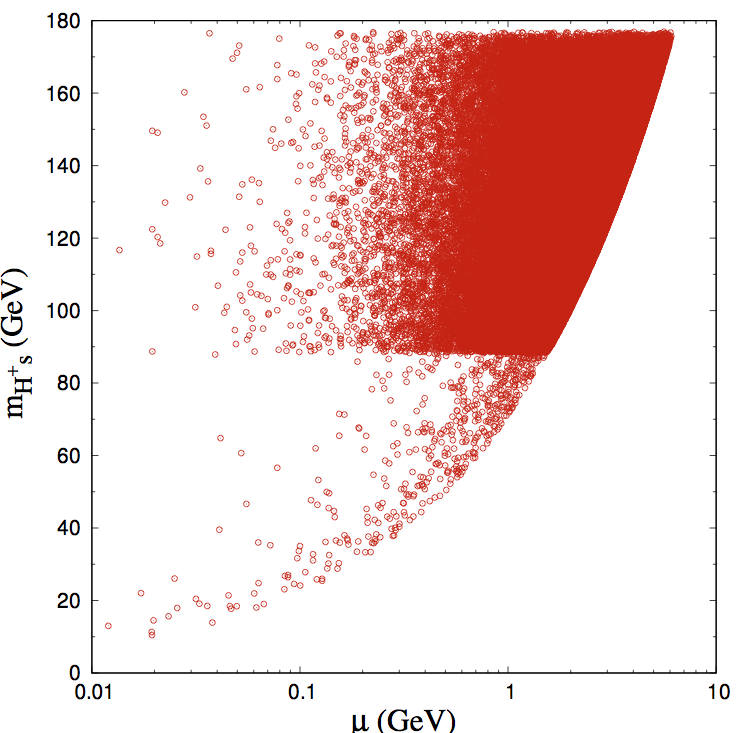}}
\caption{$m_{h^0}$ (left) and $m_{H^\pm}$ (right) dependences on $v_t$ (upper panel) and $\mu$ (lower panel). Input parameters are: $\lambda \approx 0.52$ ($m_{H^0} \approx 125$ GeV), $| \lambda_{a} | \le 1.5$, $0 \le \lambda_{b} \le 1$, $10^{-2} \le \mu \le 10$ (GeV)  and $10^{-2} \le v_t \le 3$ (GeV)}
\label{fg:fig6}
\end{figure*}

\begin{figure*}[t!]
\centering
\resizebox{0.38\textwidth}{!}{
\includegraphics{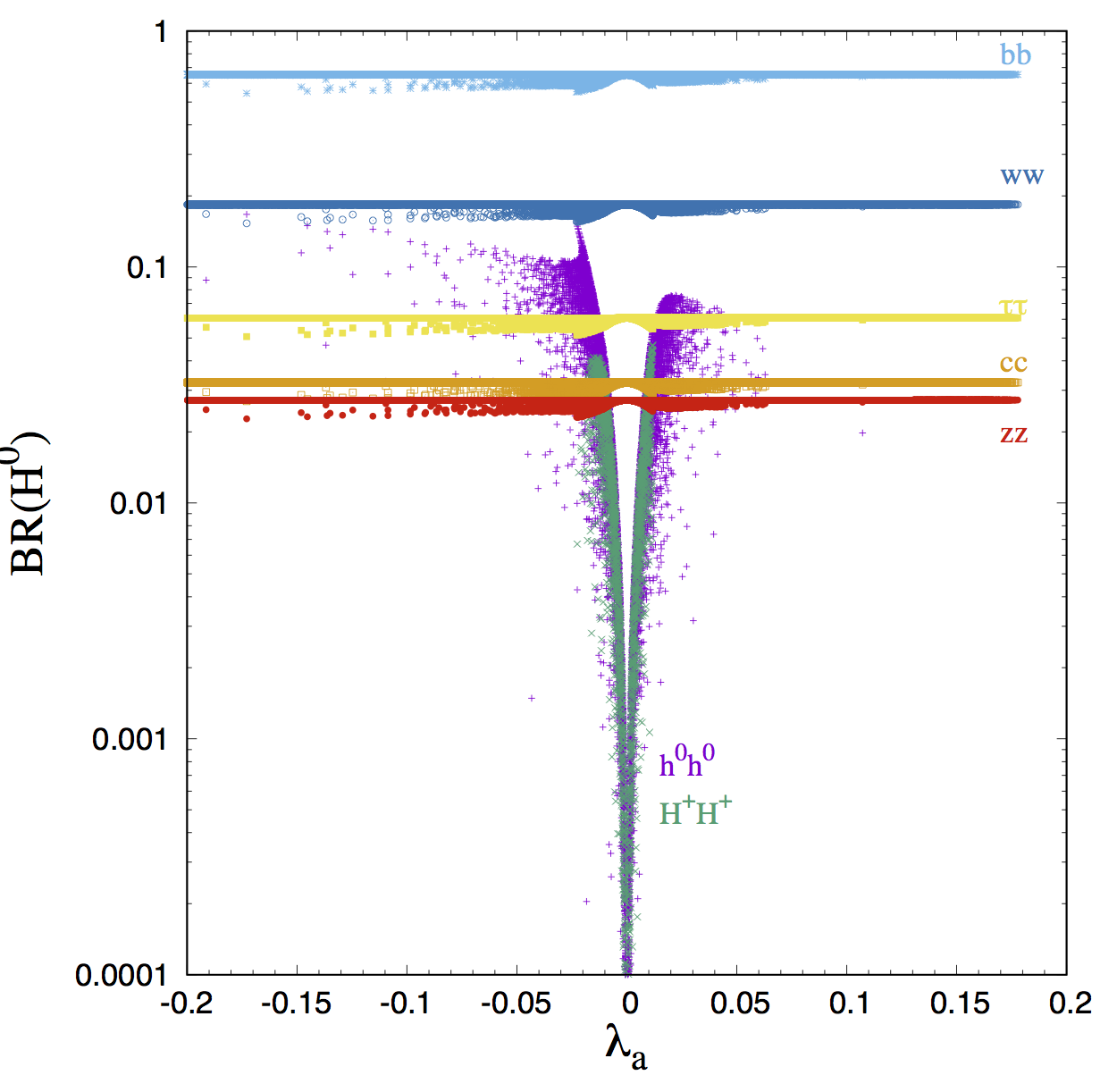}}
\caption{Branching ratio of $H^0\,\to$ $b\bar{b}$, $c\bar{c}$, $\tau^+ \tau^-$, 
$W^+ W^-$, $Z Z$, $h^0 h^0$ and $H^\pm H^\mp$ as a function of $\lambda_{a}$. Our inputs are $\lambda \approx 0.52$, 
$\lambda_b = 1$, $v_t = 1$ GeV and $0.1\le \mu \le 0.52$ (GeV) ($m_{H^0}\approx 125$ GeV).}
\label{fg:fig7}
\end{figure*}

%
The figure~\ref{fg:fig6} displays the dependence of light and charge Higgs bosons masses on $\mu$ and $v_t$ parameters when the heavier CP-even state $H^0$ is identified to the SM-like Higgs boson. At first glance, the default values of these parameters for a given region where $m_{h^0}\le\,\frac{m_{H^0}}{2}$ should not be of the same order of magnitude, indeed, to fulfil such situation, we request $v_t$ to be equal or slightly higher than $1$ GeV for a given $\mu$ below $1$ GeV. As a results, the parameter space is quite restricted offering  many new interesting features. Indeed, the charged Higgs  is very light with an upper bound on its mass about  $180$ GeV, as can been seen from  Eq.~(\ref{eq:mHpm}). Moreover, for such small values of $\mu$, the lightest CP-even state $h^0$ is mostly  dominated by a triplet component and is typically light as can be deduced from  Eqs.~(\ref{eq:mh1}- \ref{eq:sa3}).  Thus, in this scenario the LEP constraints apply to  $h^0$ Higgs.  At LEP colliders, the Higgs was searched for essentially in the channel $e^+e^- \to h^0Z \to bbZ $ in association with Z boson.  From the combined data collected by the LEP experiments, a lower limit on the Higgs mass has been established, $m_h > 114.4$ GeV, as well as a set of upper bounds on the Higgs coupling to Z boson \cite{opal,delphi}. Hence from these LEP results, one can figure out which region  of the parameter space which would be allowed (or excluded).  In HTM0 model,  the coupling of the lightest Higgs to Z boson coupling $h^0ZZ$, which is proportional to $c_\alpha=cos\alpha \approx 0$, is heavily suppressed with respect to that of the SM \cite{arhrib2014}.  Hence, the $hZ$ cross section is drastically reduced and the $h^0$ Higgs may have a mass below the $114.4$  limit, while still being in agreement with the LEP constraints.  

 It is worth to notice that, according to Eq.~(\ref{eq:lamd}), the mass of the heavier CP-even state $H^0$ matches the observed value $m_{H^0} \approx 125$ GeV, if the coupling $\lambda$ is approximately set to the value $\lambda \approx 0.52$. Such scenario offers a particularly rich 
phenomenology. Our analysis will focus on two interesting Higgs to Higgs decays, namely: $H^0 \to h^0 {h^0}^{(*)}, H^\pm {H^\pm}^{(*)}$. These invisible Higgs decay channels might become kinematically favoured with significant branching ratios for certain regions of the HTM$0$ parameter space. Indeed, again as $|s_\alpha| \approx 1, c_\alpha \approx 0$ in these regions, the $h^0h^0H^0$ and $H^\pm H^\pm H^0$ couplings reduce to,
\begin{equation}
g_{h^0 h^0 H^0} = g_{H^\pm H^\pm H^0} \simeq \lambda_{a} v_d + {\cal O}(v_t)
\label{eq:hhH-hphpH-couplings}
\end{equation}

Then, we plot in Fig.~\ref{fg:fig7} the branching ratios of the $H^0$ decays into $b\bar{b}$, $c\bar{c}$, $W^+W^-$  $ZZ$, and into the invisible decay modes $h^0h^0$ and $H^\pm H^\mp$. We clearly see that the branching ratios into $h^0h^0$ and $H^\pm H^\mp$ become dominant for non-vanishing values of $|\lambda_{a}|$, as can be seen from Eq.~(\ref{eq:hhH-hphpH-couplings}) where the corresponding couplings get substancially large values. However, once $\lambda_{a}$ approaches zero, these decay channels fade away.

\begin{figure*}[t!]
\centering
\resizebox{0.38\textwidth}{!}{
\includegraphics{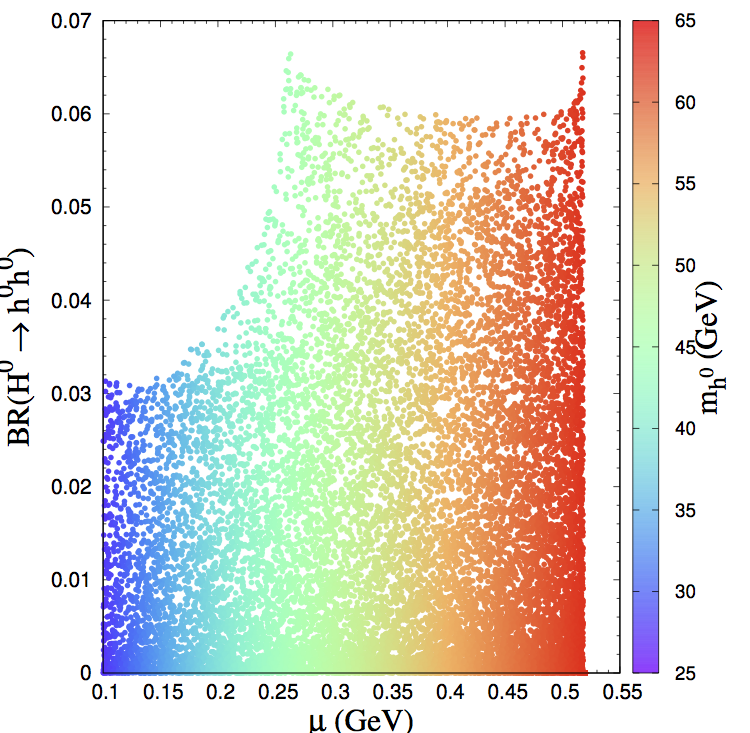}}
\resizebox{0.38\textwidth}{!}{
\includegraphics{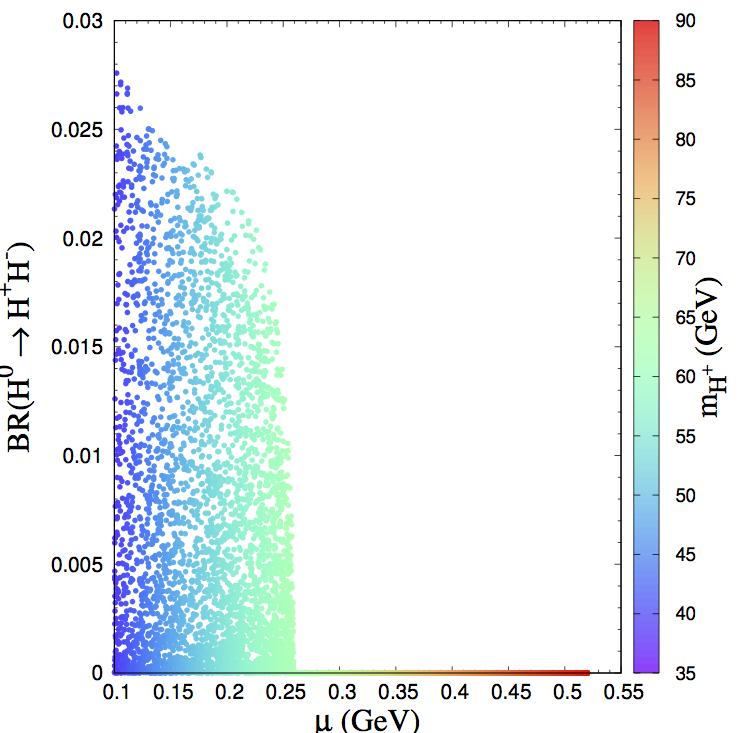}}
\caption{The branching ratios for $H^0 \to h^0h^0$ (left)  and $H^0 \to H^\pm\,H^\mp$ (right) as a function of $\mu$. The Higgs masses, $m_{h^0}$ and $m_{H^\pm}$, are considered in the ranges represented by the color codes.  Our inputs are : $\lambda \approx 0.52$, $10^{-2} \le \mu \le 0.55$ (GeV), $m_{H^0}\approx 125$ GeV, $| \lambda_{a} | \le 1.5$, $\lambda_{b} = 1$,  and $v_t = 1$ (GeV)}
\label{fg:fig8}
\end{figure*}

By the following, we fix $v_t=1$ GeV and $\lambda_b=1$, we present in 
Fig.~\ref{fg:fig8} the branching ratios for $H^0 \to h^0h^0$ and $H^0 \to H^\pm\,H^\mp$. From the left panel, we can see that decay into $h^0h^0$ gets sizeable values for values of the $\mu$ parameter larger than $0.15$ GeV ($m_{h^0}\approx\,35$ GeV), reaching up to $7\%$ when $m_{h^0}$ is around $45\sim50$ GeV. When $\mu$ becomes larger than $0.26$ GeV ($m_{h^0}\approx\,45$ GeV), this ratio decreases slightly but still remains relatively important, and never falls below $6\%$. Furthermore, for $m_{h^0} \approx 60\sim65$ GeV,  it raises to reach $7\%$ again.

\begin{figure*}[t!]
\centering
\resizebox{0.38\textwidth}{!}{
\includegraphics{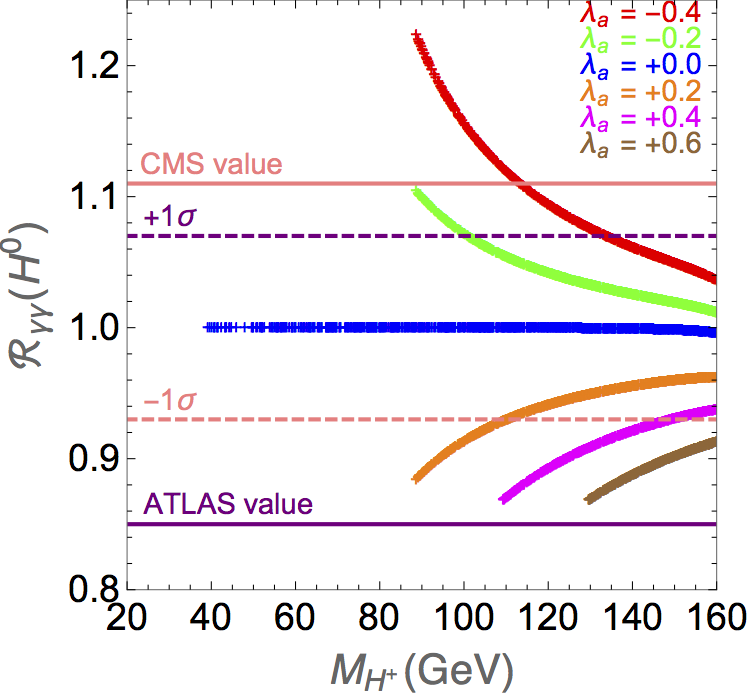}}
\resizebox{0.38\textwidth}{!}{
\includegraphics{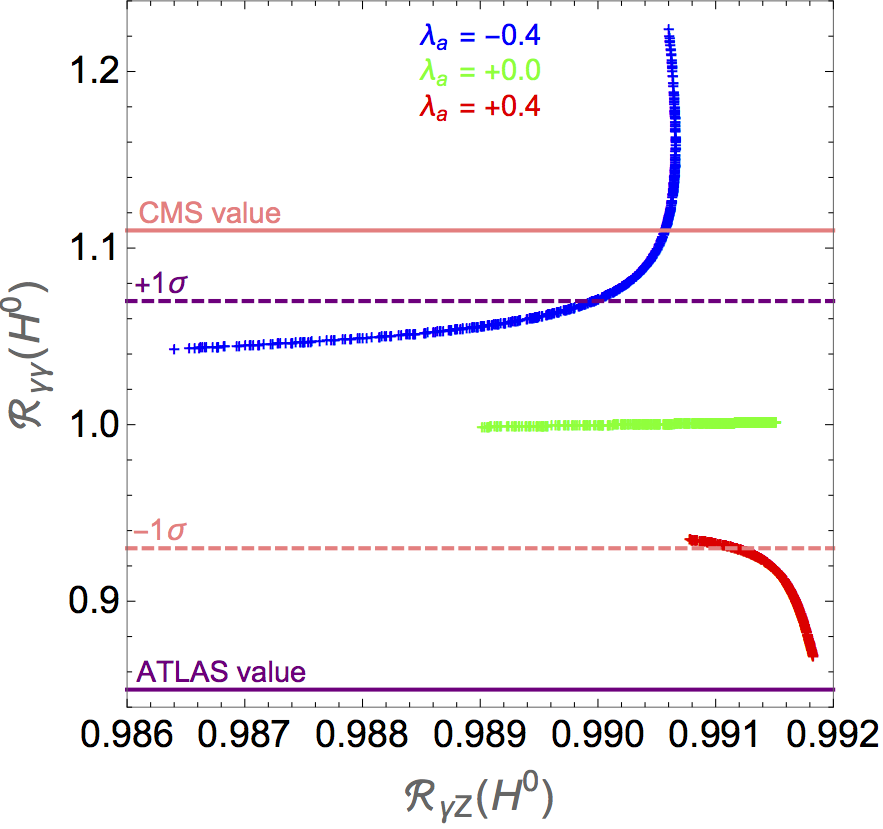}}
\caption{(Left) : $R_{\gamma\gamma}(H^0)$ as a function of $m_{H^\pm}$ for various values of $\lambda_a$. (Right) : correlation between $R_{\gamma\gamma}(H^0)$ and $R_{\gamma\,Z}(H^0)$ for various of $\lambda_a$. Our inputs are : $\lambda \approx 0.52$, $0.5 \le \mu \le 1.6$ (GeV)  ($m_{H^0}\approx 125$ GeV), $\lambda_{b} =1$ and $v_t = 1$ GeV.} 
\label{fg:fig9}
\end{figure*}

The situation is quite different for the ${\rm BR}(H^0 \to H^\pm H^\pm)$ as illustrated in the right panel of Fig.~\ref{fg:fig8}. This ratio tends to its maximal value, $\approx2.7\%$,  for very tiny $\mu$ about $\approx 0.1$ GeV, corresponding to small values of $m_{H^\pm}\approx 39$ GeV, and decreases inversely when $\mu$ increases up to the value $\mu\approx\,0.26$ GeV. In contrast to the decay into $h^0h^0$, beyond this value, the branching ratio is almost vanishing.

From the left side of Fig.~\ref{fg:fig9}, the ratio $R_{\gamma\gamma}(H^0)$ reaches its SM-like value for $\lambda_a\approx\,0$ and for the charged Higgs mass in the range $40\sim160$ GeV, while an excess up to $20\%$ can be achieved for negative values of $\lambda_a$. If ATLAS/CMS exclusions data at 1$\sigma$, is taken into account, then this excess is largely reduced to less than 10$\%$. As a byproduct, this analysis sets up a lower limit on the $m_{H^\pm}$ of order $\sim115$ GeV (for $\lambda_a=-0.2$). In addition, $R_{\gamma\gamma}(H^0)$ remains below it SM value when $\lambda_a > 0$, even for $m_{H^\pm}$ above this lower value. At last, we study correlation of $R_{\gamma\,\gamma}(H^0)$ with $R_{\gamma\,Z}(H^0)$ in this scenario. Unlike the $h^0$ SM-like case, one can see from the right panel of Fig.~\ref{fg:fig9} that these observables are correlated for $\lambda_a<0$ or anti-correlated for $\lambda_a>0$ with a predicted charged Higgs mass in the range $[130\sim160]$ or $[110\sim160]$ GeV respectively.

\subsection{Degenerate case : $m_{H^0} \approx m_{h^0} \approx\,125$ GeV \label{sec:degenerate}}
In this subsection, we consider the CP-even neutral Higgs bosons $h^0$ and $H^0$ with nearly degenerate mass. This scenario has recently attracted attention and been taken seriously  in many SM extensions \cite{chabab14,gunion-kraml-13,ferreira2013,drozd2013}.  Here we would like to ask to what extent this survives in HTM$0$ in light of LHC data at $13$ TeV.  In other words, we probe the region of the parameter space where the twin Higgs decays into diphoton Higgs with branching ratio (or signal strength $R_{\gamma\gamma}$) consistent with ATLAS and CMS data. A first analysis has been performed in \cite{wang2014}. This analysis used an intriguing  and unjustified hypothesis considering the charged Higgs mass equals to the neutral ones. In this model, this possibility is excluded by theoretical constraint as we will show shortly.  But first, we will demonstrate that the parameter space is restricted further by an additional constraint, induced by the Higgs mass degeneracy, and leading to a severe control of the potential parameters.      

The two eigenvalues $m_{\pm}$ (with $m_{-}=m_{h^0}^2<m_{+}=m_{H^0}^2$), representing the squared masses of  $h^0$ and $H^0$, are :
\begin{equation}
m_{\pm} = \frac{A+C  \pm \sqrt{(A-C)^2 + 4 B^2}}{2}.
\label{dgnr1}
\end{equation}
Then
\begin{eqnarray}
m_+ - m_- &=& (m_{H^0} - m_{h^0}) (m_{H^0} + m_{h^0}) \nonumber\\
& \approx & (m_{H^0} - m_{h^0}) 2 M_{ex} = 2 M_{ex} \Delta M. \nonumber
\end{eqnarray}

where $\Delta M$, the difference of masses between the two neutral Higgs $H^0$ and $h^0$ is set to
about $1$~GeV, corresponding to the detector inability to resolve two nearly Higgs signals, and $M_{ex}$ is the experimental Higgs boson mass $\approx 125$ GeV. Taking into account these considerations one gets $\displaystyle{ \sqrt{(A-C)^2 + 4 B^2}} \leq 2 M_{ex} \Delta M$, that obviously leads to two constraints: $|B|\le M_{ex} \Delta M$ and $|A-C|\leq 2\,M_{ex} \Delta M$.

\begin{figure*}[t!]
\centering
\resizebox{0.38\textwidth}{!}{
\includegraphics{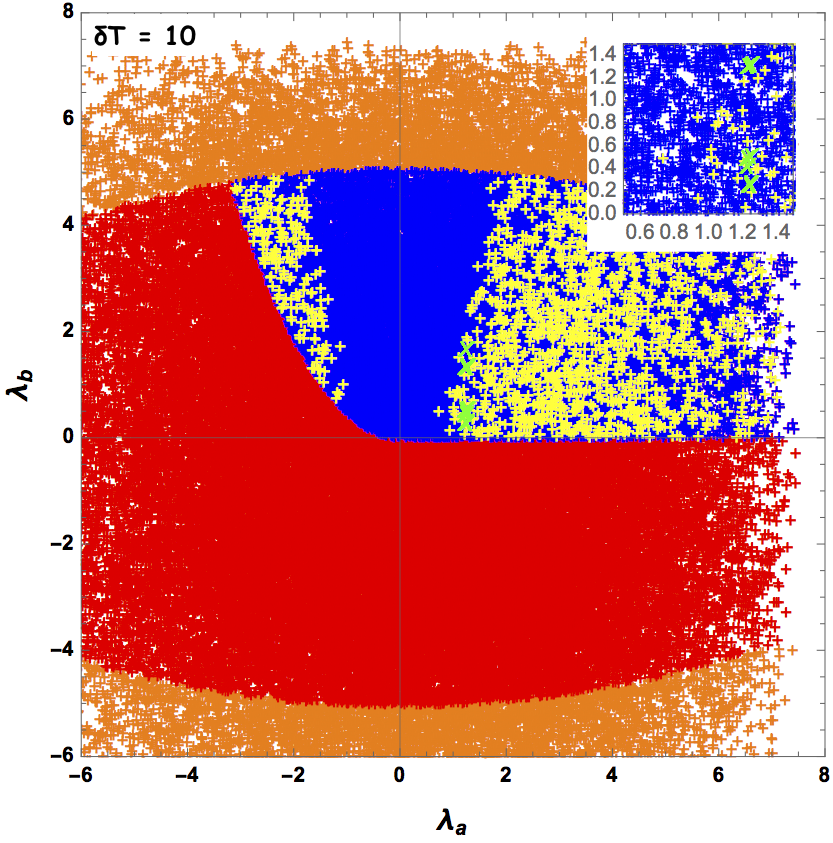}}
\caption{The allowed regions in ($\lambda_{a},\lambda_{b}$) for $\delta T = 10$ in the degenerate case. Color codes are as follows, \textcolor{orange}{Orange} : Excluded by Unitarity constraints. \textcolor{red}{Red} : Excluded by Unitarity+BFB constraints. \textcolor{blue}{Blue} : Excluded by Unitarity+BFB \& $\displaystyle{\frac{\mu}{\lambda}   \approx  4\,v_t}$ constraints. \textcolor{yellow}{Yellow} : Excluded by Unitarity+BFB \& $T_d \approx \delta T$ $\land$ $T_t \approx \delta T$ \& $\displaystyle{\frac{\mu}{\lambda} \approx  4\,v_t}$ constraints.
That shows only the \textcolor{green}{Green} area obeys  ALL constraints. Our inputs are $\lambda = 0.52$, 
$-5 \le \lambda_1, \lambda_2, \lambda_3, \lambda_4, \lambda_5 \le 5$, $10^{-3} \le v_t \le 3$ (GeV) and $10^{-3} \le \mu \le 5$ (GeV).}
\label{fg:fig10}
\end{figure*}

The first constraint reads as:
\begin{equation}
|2\,\lambda_a\,v_t - \mu | \leq 2\,\sqrt{2}\, \frac{M_{ex} \Delta M }{v_d}, 
\label{dgnr2}
\end{equation}
while, for small ratio of the two vevs $\frac{v_t}{v_d}$, the second constraint reduces to,
\begin{equation}
|4\,\lambda\,v_t - \mu | \leq \frac{8\,v_t}{v_d}\,\frac{2\,M_{ex} \Delta M }{v_d}, 
\label{dgnr3}
\end{equation}
Since the ratio $\frac{2\,M_{ex} \Delta M }{v_d}$ is about 1 GeV, these two relations simplify to $|2\,\lambda_a\,v_t - \mu| \leq \sqrt{2}$ GeV and $\displaystyle{\frac{\mu}{\lambda}   \approx  4\,v_t}$, providing strict bounds to the three potential parameters $\mu$, $\lambda$ and $\lambda_a$, hence severely reducing the allowed regions in the parameter space, as it is illustrated in Fig.\ref{fg:fig10}. This feature has a dramatic effect on the discrepancy between the neutral and charged Higgs masses as can be seen from Fig.\ref{fg:fig2}. In such figure, the Higgs bosons masse behaviours are plotted as a function of the $\mu$ parameter; 
these values satisfy the above resulting relation in the degenerate case. The seemingly constant $m_{h^0}^2$ for $\mu > \mu_c$ and constant $m_{H^0}^2$ for $\mu < \mu_c$ are clearly achieved around the critical value $\mu_c\approx\,2.1$ GeV. Contrary to what one might think, if we take the Higgs bosons masses as inputs \cite{wang2014}, such a situation matches a splitting between the charged Higgs boson mass and the $\mathcal{H}$ ($=h^0=H^0$) degenerate state mass in the range of $\Delta\,m\,=\,m_{H^\pm}-m_{\mathcal{H}}\approx\,51$ GeV.

\begin{figure*}[t!]
\centering
\resizebox{0.38\textwidth}{!}{
\includegraphics{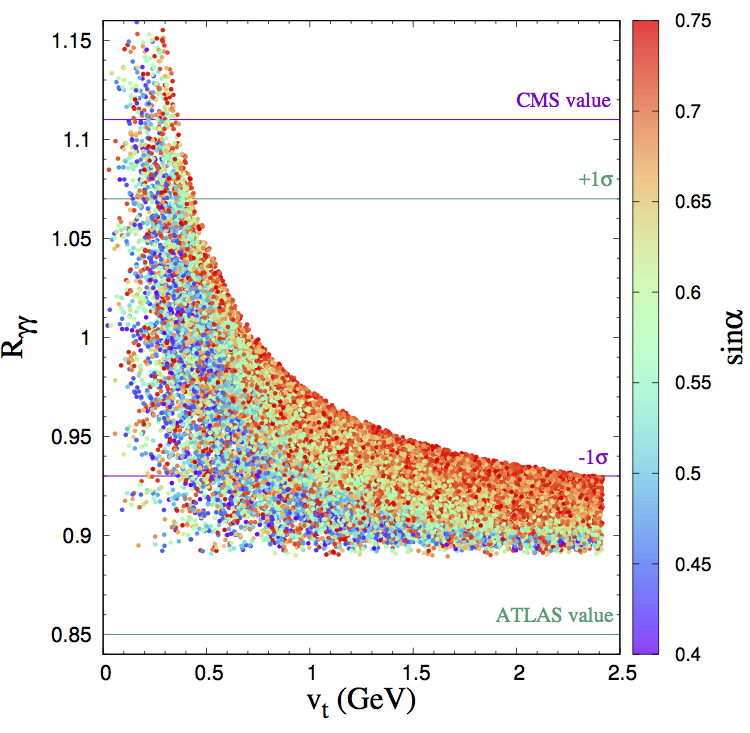}}
\resizebox{0.38\textwidth}{!}{
\includegraphics{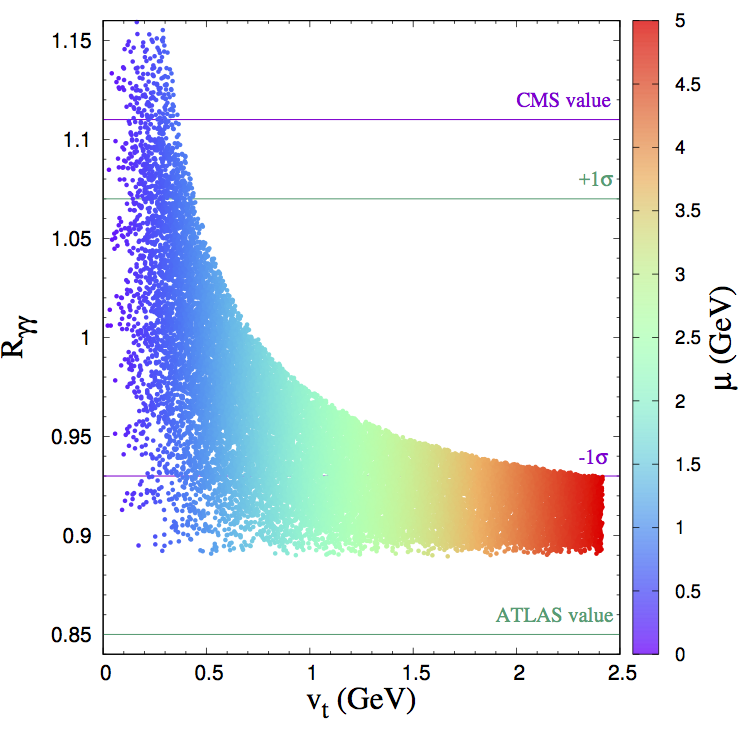}}
\caption{Left: Scatter plot for $\sin\alpha$ in the plan $(R_{\gamma\gamma},v_t)$ with $10^{-3} \le \mu \le 5$ (GeV). Right: $R_{\gamma\gamma}$ as function of $v_t$, where the palette shows the size of $\mu$.}
\label{fg:fig11}
\end{figure*}

Hereafter we define the diphoton signal strength $R_{\gamma\gamma}$ by the following quantity,
\begin{equation}
R_{\gamma\gamma}=R_{\gamma\gamma}(h^0)+R_{\gamma\gamma}(H^0) 
\label{eq:RgagaTOT}
\end{equation}
and similarly $R_{\gamma\,Z}$ is introduced.
In this scenario, the charged Higgs boson loops are included with the $g_{\mathcal{H}ww},  g_{\mathcal{H}\bar{f}f}$ couplings given by Table.~\ref{table_couplings}.

Fig.~\ref{fg:fig11} illustrates the HTM0 degenerate case effect on $R_{\gamma\gamma}$. Similarly to the previous scenarios, we fix $\lambda\sim0.518$ and scan over $\lambda_{a}$, $\lambda_{b}$, $\mu$ and $v_t$, with the Higgs masses given by Eqs.~\ref{dgnr1}, \ref{dgnr2} and \ref{dgnr3}. In the left panel, we show the scatter plot for the mixing angle $\alpha$ in the $(R_{\gamma\gamma},v_t)$ plane. Again we see that small but no zero values below $0.5$ are favoured for the triplet {\it vev} $v_t$ to achieve the standard limit, corresponding to $\sin\alpha\sim0.55-0.65$. Equally, as set out from its dependence on $v_t$ in this scenario, the $\mu$ parameter takes a tiny values. In the right panel, we show the variation of $R_{\gamma\gamma}$ a function of $\mu$ and $v_t$ within $1\sigma$ of ATLAS/CMS measurements.

\begin{figure*}[t!]
\centering
\resizebox{0.38\textwidth}{!}{
\includegraphics{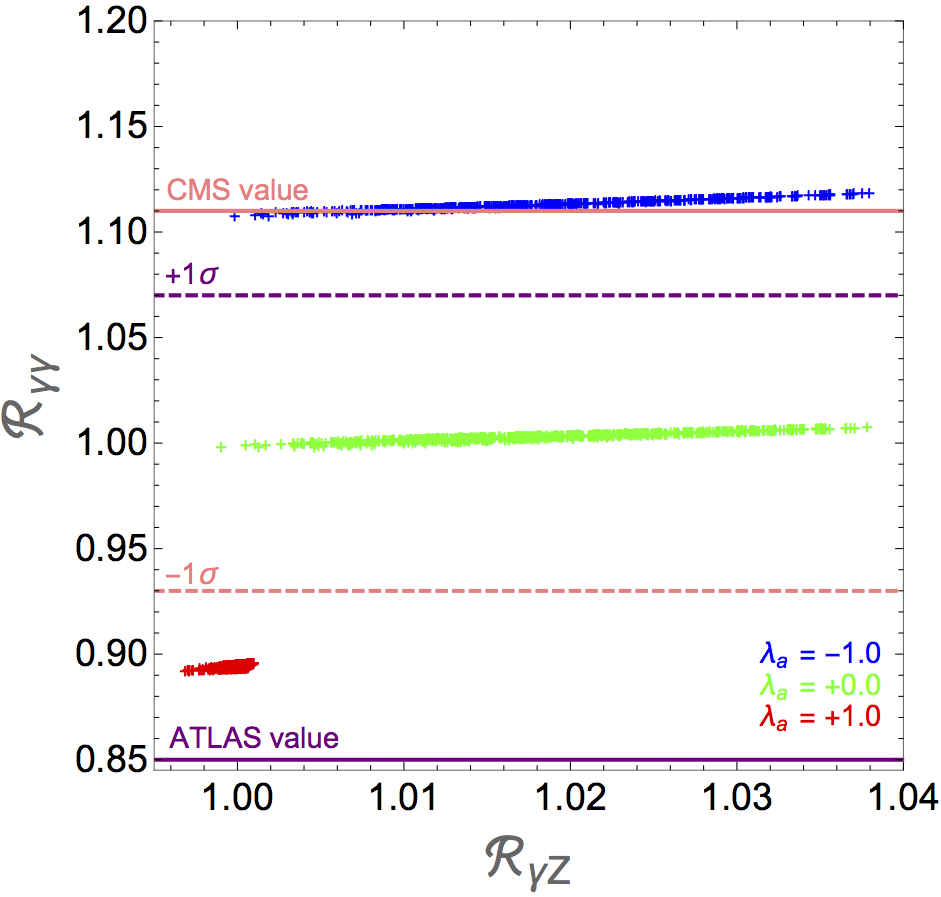}}
\caption{$R_{\gamma\gamma}$ and $R_{\gamma\,Z}$ correlation in the degenerate case 
for various $\lambda_a$. Inputs are the same as in Fig.\ref{fg:fig11}, except for $\lambda_a$.}
\label{fg:fig12}
\end{figure*}

Finally, we display in Fig.~\ref{fg:fig12}, we have plotted $R_{\gamma\gamma}$ versus $R_{\gamma\,Z}$ in mass degenerate scenario for various values of $\lambda_a$. From this plot one can see that the correlation is always positive whatever the value of $\lambda_a$. We also note that no noticeable enhancement can be achieved, since most part of the parameter space is drastically constrained by a constant charged Higgs mass at about $m_{H^\pm}\sim\,176$ GeV, as shown form Fig.~\ref{fg:fig2}, which concurs with the results predicted in \cite{chabab14}.

\section{Conclusion}
In this paper, we have discussed some features of the Higgs triplet model with null hypercharge (HTM0),  an extension of the SM with a larger scalar sector.  First, we have shown that the parameter space of HTM0 generally constrained by unitarity and boundedness from below, is severely reduced when the modified Veltman conditions are imposed.  Then, we have investigated some Higgs decays, including Higgs to Higgs decays,  in light of LHC data, either when $h^0$ is the SM-like Higgs or when the heaviest neutral Higgs $H^0$ is identified to the  $125$  observed GeV Higgs. In addition, we have analysed the degenerate scenario and shown that LHC signal strengths favours a light charged Higgs mass about $176\sim178$ GeV. Finally,  we have pointed out some discrepancies with previous analysis, regarding the correlations between the diphoton Higgs decay mode and $H \to  Z \gamma$ mode.

\section*{Acknowledgment}
The authors would like to thank G. Moultaka for useful discussions.
MC and LR would like to thank  LUPM laboratory at Montpellier University for hospitality. 
This work is supported in part by the Moroccan Ministry of Higher Education and Scientific Research under 
contract N$^{\circ}$PPR/2015/6, and by the GDRI--P2IM: Physique de l'infiniment petit et l'infiniment grand.

\section*{Appendix A : Unitarity constraints}
\label{sec:appendix:a}
By exploring the HTM$0$ model, we can show that the full set of $2$-body scalar scattering processes leads to
a $19\times19$ S-matrix with $5$ block of submatrices corresponding to mutually
 unmixed sets of channels with definite charge and CP states. Hence one gets the following submatrix dimensions,  
structured in terms of net electric charge in the initial/final states:   
${S}^{(1)}(4 \times 4)$, ${ S}^{(2)}(5 \times 5)$ and ${ S}^{(3)}(1 \times 1)$, corresponding to 
$0$-charge channels, ${ S}^{(4)}(6 \times 6)$ for the $1$-charge channels,
and  ${S}^{(5)}(3 \times 3)$ corresponding to the $2$-charge channel.   \\

In principle, by using the unitarity equation,  one can derive the unitarity constraint on each component of the
S-matrix. Thus the usual unitarity bound on partial wave amplitudes would apply to the eigenvalues of the submatrices,   encoding indirectly the bounds on all the components $\tilde{T}^{(n)}$ of the T-matrix, defined as ${\cal M}_n \equiv i \tilde{T}^{(n)}$, with $n=1, \cdots, 5$.  \\

We present hereafter the resulting submatrices whose entries correspond  to the quartic couplings that mediate the $2 \to 2$ scalar processes.
By writing the neutral components in the fields as : $\phi^0=\frac{1}{\sqrt{2}}(v_d+h_1+i\,z_1)$ and $\delta^0=v_t+h_2$
, the first submatrix ${\cal M}_1$  corresponds to scattering whose
initial and final states are one of the following:
$(\phi^+\delta^-$,$\delta^+\phi^-$, $h_2 z_1
$,  $h_1 h_2)$. We have to write out the full matrix,
one finds,
\begin{eqnarray}
{\cal M}_1 = \left(
\begin{array}{cccc}
\displaystyle \lambda_a & 0 & \displaystyle 0 & \displaystyle 0   \\
 0 & \displaystyle \lambda_a & \displaystyle 0 & \displaystyle 0  \\
 0 & \displaystyle 0 & \displaystyle \lambda_a & \displaystyle 0   \\
  0 & \displaystyle 0 & \displaystyle 0 & \displaystyle \lambda_a  \\
\end{array}\right)
\end{eqnarray}

The second submatrix ${\cal M}_2$ corresponds to scattering with one of the following
initial and final states:
$(\phi^+\phi^-$, $\delta^+\delta^-$, $\frac{h_1 h_1}{\sqrt{2}}$, 
$\frac{h_2 h_2}{\sqrt{2}}$,
$\frac{z_1 z_1}{\sqrt{2}})$, where the $\sqrt{2}$ accounts for
identical particle statistics. From a straightforward calculation, one finds that ${\cal M}_2$ reads as:
\begin{eqnarray}
{\cal M}_2 = \left(
\begin{array}{ccccc}
\displaystyle \lambda & \displaystyle \lambda_a & \displaystyle \frac{\lambda}{2\sqrt{2}} & \displaystyle \frac{\lambda_a}{\sqrt{2}} & \displaystyle \frac{\lambda}{2\sqrt{2}} \\
\displaystyle \lambda_a & \displaystyle 4\lambda_b & \displaystyle \frac{\lambda_a}{\sqrt{2}} & \displaystyle \sqrt{2}\,\lambda_b & \displaystyle \frac{\lambda_a}{\sqrt{2}} \\
\displaystyle \frac{\lambda}{2\sqrt{2}} & \displaystyle \frac{\lambda_a}{\sqrt{2}} & \displaystyle \frac{3\lambda}{4} & \displaystyle \frac{\lambda_a}{2} & \displaystyle \frac{\lambda}{4}\\
\displaystyle \frac{\lambda_a}{\sqrt{2}} & \displaystyle \sqrt{2}\,\lambda_b & \displaystyle \frac{\lambda_a}{2} & \displaystyle 3\,\lambda_b & \displaystyle \frac{\lambda_a}{2} \\
\displaystyle \frac{\lambda}{2\sqrt{2}} & \displaystyle \frac{\lambda_a}{\sqrt{2}} & \displaystyle \frac{\lambda}{4} & \displaystyle \frac{\lambda_a}{2}& \displaystyle \frac{3\lambda}{4} \\
\end{array}
\right)
\end{eqnarray}

\noindent
 Despite its apparently complicated structure, the seven eigenvalues of ${\cal M}_2$ can be easily determined. At last, for the 0-charge $2\to2$ processus,  there is just one state $h_1 z_1$ leading to ${\cal M}_3 = \lambda/2$

On the other hand, the $1$-charge channels occur for two-by-two body scattering
between the charged states $ (h_1 \phi^+ $,  $z_1 \phi^+ $, $h_2 \phi^+ $,
$h_1 \delta^+$, $z_1 \delta^+$, $h_2 \delta^+)$.
The 6$\times$6 submatrix ${\cal M}_4$ obtained from the above
scattering processes is given by:
\begin{eqnarray}
{\cal M}_4=\left(
\begin{array}{cccccc}
\displaystyle \frac{\lambda}{2} & 0 & 0 & 0 & 0 & 0 \\
\displaystyle 0 & \displaystyle \frac{\lambda}{2} & 0 & 0 & 0 & 0 \\
\displaystyle 0 & 0 & \displaystyle \lambda_a & 0 & 0 & 0 \\
\displaystyle 0 & 0 & 0 & \displaystyle \lambda_a & 0 & 0 \\
\displaystyle 0 & 0 & 0 & 0 & \displaystyle \lambda_a & 0 \\
\displaystyle 0 & 0 & 0 & 0 & 0 & \displaystyle 2\lambda_b \\
\end{array}
\right)
\end{eqnarray}

while the fifth submatrix ${\cal M}_5$ corresponds to scattering with
initial and final states being one of the following $3$ sates:
$(\frac{\phi^+\phi^+}{\sqrt{2}}$,$\frac{\delta^+\delta^+}{\sqrt{2}}$,$\delta^+\phi^+)$. It reads,

\begin{eqnarray}
{\cal M}_5=\left(
\begin{array}{ccc}
\displaystyle \frac{\lambda}{2} & 0 & 0  \\
\displaystyle 0 & \displaystyle{2\lambda_b} & 0  \\
\displaystyle 0 & 0 & \displaystyle{\lambda_a} \\
\end{array}
\right)
\end{eqnarray}

From the usual expansion in terms of partial-wave amplitudes $a_J$, we write, 
following our notations, 
\begin{equation}
{\cal M}^{(k f)} = i \tilde{T}_{k f} = 16 i \pi \sum_{J\ge0} ( 2 J + 1) \, a^{(k f)}_J(s) \, P_J(\cos \theta)
\end{equation}
where $k$ and $f$ run
over all possible initial and final states of the above $19$-state basis and the $P_J$'s are the Legendre polynomials. 
Since we only consider the leading high energy contributions for each channel, all the partial waves with $J\neq 0$ vanish,except one: 
\begin{equation} a^{(k f)}_0 = - \frac{i}{16 \pi} {\cal M}^{(k f)} \label{eq:Unitconst} \end{equation}
 The S-matrix unitarity constraint for elastic scattering, $|a^{(kk)}_0| \le 1$ or alternatively $| Re(a^{(kk)}_0)| \le \frac{1}{2}$, translates through Eq.~(\ref{eq:Unitconst}) directly to all the eigenvalues of the submatrices we determined above.

\section*{Appendix B : Feynman Rules for tadpoles}
\label{sec:appendix:b}
In this appendix, we list the couplings used to calculate the tadpoles of the two neutral CP-even Higgs $h^0$ and $H^0$ as explained in \cite{chabab16}.

We note $c_{F_i \bar {F_i}}$ ($C_{F_i \bar {F_i} }$) the couplings to the Higgs $h^0$ ($H^0$) where $F_i$ stands for any quantum field of the HTM0: scalar and vectorial bosons, fermions, Goldstone fields $G_i$ and Faddeev-Popov ghost fields $ \eta_i$. Because the field $F_i$ fixes the propagator, we also list the values $t_i$ ($T_i$) of the loop due to the propagator of the $F_i$ particle which gain a factor $2$ in case of charged fields, and the symmetry factor $s_i$.

\begin{eqnarray}
&& c_1 \equiv c_{h_0h_0} = -\frac{3 i}{2} (\lambda v_d c_\alpha^3 + 2 \lambda_a v_d c_\alpha s_\alpha^2 + 4 \lambda_b v_t s_\alpha^3 +\nonumber\\
&&\hspace{2.5cm} (-\mu + 2\,\lambda_a v_t) c_\alpha^2 s_\alpha), \nonumber\\
&& C_1 \equiv C_{H_0 H_0} = \frac{3 i}{2} (\lambda v_d s_\alpha^3 + 2 \lambda_a v_d s_\alpha c_\alpha^2 - 
4 \lambda_b v_t c_\alpha^3 - \nonumber\\
&& \hspace{2.5cm} (-\mu + 2\,\lambda_a v_t) s_\alpha^2 c_\alpha), \nonumber\\
&& t_1 = i A_0(m_{h_0}^2),\nonumber\\
&& T_1 = i A_0(m_{H_0}^2), \nonumber\\
&& s_1= \frac{1}{2},
\end{eqnarray} 
\begin{eqnarray}
&& c_2 \equiv c_{G_0 G_0} = -\frac{ i}{2} (-\mu s_\alpha + \lambda v_d c_\alpha + 2 \lambda_a s_\alpha v_t),\nonumber\\
&& C_2 \equiv C_{G_0 G_0} = +\frac{ i}{2} (\mu c_\alpha + \lambda v_d s_\alpha - 2 \lambda_a c_\alpha v_t),\nonumber\\
&& t_2 = T_2 = i A_0(\xi_Z m_Z^2),\nonumber\\
&& s_2= \frac{1}{2},
\end{eqnarray}
\begin{eqnarray}
&& c_3 \equiv c_{G_+ G_-} = -\frac{i}{2}(2\,\mu c_\alpha c_{{\theta_{\pm}}} s_{{\theta_{\pm}}} + (\mu s_{\alpha} + 
\lambda v_d c_\alpha + 2 \lambda_a v_t s_\alpha) c_{\theta_{\pm}}^2 \nonumber\\
&&\hspace{2.4cm}+ 2 (\lambda v_d c_\alpha + 2 \lambda_b v_t s_\alpha) s_{\theta_{\pm}}^2), \nonumber\\
&& C_3 \equiv C_{G_+ G_-} = -\frac{i}{2}(-(2\mu c_{{\theta_{\pm}}} s_{{\theta_{\pm}}} + \lambda v_d c_{\theta_{\pm}}^2 + 
2\lambda_a v_d s_{\theta_{\pm}}^2)\,s_\alpha \nonumber\\
&&\hspace{2.4cm}+ (4 \lambda_b s_{{\theta_{\pm}}}^2 v_t + 2 \lambda_a v_t c_{\theta_{\pm}}^2 + \mu c_{\theta_{\pm}}^2)\,c_\alpha),\nonumber\\ 
&& t_3  = T_3 = 2 \times i A_0(\xi_W m_W^2),\nonumber\\
&& s_3= \frac{1}{2},\nonumber\\
\end{eqnarray}
\begin{eqnarray}
&& c_4 \equiv c_{H_0H_0} = -\frac{i}{2} (2 \lambda_a c_\alpha^3 v_d + 
(3\lambda - 4\lambda_a) c_\alpha s_\alpha^2 v_d - (\mu- \nonumber\\
&&\hspace{2.4cm}2 \lambda_a v_t) s_\alpha^3+ 2 c_\alpha^2 s_\alpha (\mu -2 (\lambda_a-3\lambda_b) v_t) )\nonumber\\
&& C_4 \equiv C_{h_0 h_0} = -\frac{i}{2} (-2 \lambda_a s_\alpha^3 v_d - 
(3\lambda - 4\lambda_a) s_\alpha c_\alpha^2 v_d - (\mu- \nonumber\\
&&\hspace{2.4cm} 2 \lambda_a v_t) c_\alpha^3+ 2 s_\alpha^2 c_\alpha (\mu -2 (\lambda_a-3\lambda_b) v_t) )\nonumber\\
&& t_4= i A_0(m_{H_0}^2), \nonumber\\
&& T_4= i A_0(m_{h_0}^2), \nonumber\\
&& s_4= \frac{1}{2},
\end{eqnarray}
\begin{eqnarray}
&& c_5 \equiv c_{H_+ H_-} =  -\frac{i}{2}( (-2\mu c_{{\theta_{\pm}}} s_{{\theta_{\pm}}} + 2 \lambda_a c_{{\theta_{\pm}}}^2 v_d + \lambda s_{{\theta_{\pm}}}^2 v_d) c_\alpha \nonumber\\ 
&&\hspace{2.4cm} + (4 \lambda_b c_{{\theta_{\pm}}}^2 v_t + (\mu +2 \lambda_a v_t) s_{{\theta_{\pm}}}^2) s_\alpha) \nonumber\\
&& C_5 \equiv C_{H_+ H_-} =-\frac{i}{2}( 2 \mu c_{{\theta_{\pm}}} s_{{\theta_{\pm}}} s_{\alpha} + 
( \mu c_\alpha - \lambda v_d s_\alpha + \nonumber\\ 
&&\hspace{2.4cm} 2 \lambda_a v_t c_\alpha) s_{\theta_{\pm}}^2 + 2 ( - \lambda_a v_d s_\alpha +  2 \lambda_b v_t c_\alpha) c_{\theta_{\pm}}^2 ) \nonumber\\
&& t_5 = T_5 = 2 \times i A_0(m_{H_{\pm}}^2),\nonumber\\
&& s_5= \frac{1}{2},
 \end{eqnarray} 
\begin{eqnarray}
&& c_6 \equiv c_{ZZ} = i e m_W c_\alpha c_{{\theta_{\pm}}}/(c_w^2 s_w),\nonumber\\
&& C_6 \equiv C_{ZZ} = -i e m_W c_{{\theta_{\pm}}} s_\alpha/(c_w^2 s_w),\nonumber\\
&& t_6 = T_6 = -i((n-1) A_0(m_Z^2) + \xi_Z A_0(\xi_Z m_Z^2),\nonumber\\
&& s_6= \frac{1}{2},
\end{eqnarray}
\begin{eqnarray}
&& c_7 \equiv c_{W_+ W_-} = i e m_W (c_\alpha c_{{\theta_{\pm}}} - 2 s_\alpha s_{{\theta_{\pm}}})/s_w,\nonumber\\
&& C_7 \equiv C_{W_+ W_-} =-i e m_W (c_{{\theta_{\pm}}} s_\alpha - 2 c_\alpha s_{{\theta_{\pm}}})/s_w,\nonumber\\
&& t_7 = T_7 = 2 \times (-i((n-1) A_0(m_W^2) + \xi_W A_0(\xi_W m_W^2)),\nonumber\\
&& s_7 = \frac{1}{2},
\end{eqnarray}

\begin{eqnarray}
&& c_{8} \equiv c_{f \bar f}  = \frac{-i}{2} e (c_\alpha/c_{{\theta_{\pm}}}) m_f/(m_W s_w),\nonumber\\
&& C_{8} \equiv C_{f \bar f}  = \frac{i}{2}  e (s_\alpha/c_{{\theta_{\pm}}}) m_f/ m_W s_w),\nonumber\\
&& t_{8} = T_{8} = i m_f A_0(m_f^2) Tr(I_n),\nonumber\\
&& s_{8}= 1,
\end{eqnarray}
\begin{eqnarray}
&& c_{9} \equiv c_{{\eta_Z} \bar{\eta_Z}}=  \frac{-i}{2} e m_W (c_\alpha c_{{\theta_{\pm}}}) \xi_Z)/(c_w^2 s_w),\nonumber\\
&& C_{9} \equiv C_{{\eta_Z} \bar{\eta_Z}}=  \frac{ i}{2} e m_W (c_{{\theta_{\pm}}} s_\alpha) \xi_Z/(c_w^2 s_w),\nonumber\\
&& t_{9} = T_{9} = i A_0(\xi_Z m_Z^2),\nonumber\\
&& s_{9}= 1,
\end{eqnarray}
\begin{eqnarray}
&& c_{10} \equiv c_{\eta_{\pm} \bar{\eta_{\pm}}} = \frac{-i}{2} e m_W (c_\alpha c_{{\theta_{\pm}}} - 2 s_\alpha s_{{\theta_{\pm}}}) \xi_W/s_w,\nonumber\\
&& C_{10} \equiv C_{\eta_{\pm} \bar{\eta_{\pm}}} = \frac{ i}{2} e m_W (c_{{\theta_{\pm}}} s_\alpha - 2 c_\alpha s_{{\theta_{\pm}}})\xi_W/s_w,\nonumber\\ 
&& t_{10} = T_{10} = 2 \times i A_0(\xi_W m_W^2),\nonumber\\
&& s_{10}= 1,
\end{eqnarray}

\section*{Appendix C : Higgs signal strengths}
\label{sec:appendix:c}
Here we collect the Higgs signal strength measurements corresponding to various Higgs boson  production modes and Higgs decay channels.
\begin{table}[htb!]
\renewcommand{\arraystretch}{1.5}
\begin{center}
\begin{tabular}{cccc}
\hline \hline
\centering 
\multirow{2}{*}{\shortstack{Decay \\ channel}}  & \multirow{2}{*}{\shortstack{Production \\ Mode}} & \multirow{2}{*}{ATLAS} &  \multirow{2}{*}{CMS}  \\ 
& & & \\ \hline \hline
\multirow{2}{*}{$\gamma \gamma$}  & $ ggF$ & $\rm 0.62^{+0.30}_{-0.29} $\cite{ATLAS-CONF-2016-081} & $\rm 0.77^{+0.25}_{-0.23} $\cite{CMS-PAS-HIG-16-020} \\ 
& $VBF$ & $\rm 2.25^{+0.75}_{-0.75} $\cite{ATLAS-CONF-2016-081} & $\rm 1.61^{+0.90}_{-0.80} $\cite{CMS-PAS-HIG-16-020}
\\ \hline

\multirow{2}{*}{$ ZZ $} & $ ggF$ & $\rm 1.34^{+0.39}_{-0.33} $\cite{ATLAS-CONF-2016-081} & $\rm 0.96^{+0.40}_{-0.33} $\cite{CMS-PAS-HIG-16-033} \\
& $ VBF$ & $\rm 3.8^{+2.8}_{-2.2} $\cite{ATLAS-CONF-2016-081} & $\rm 0.67^{+1.61}_{-0.67}$\cite{CMS-PAS-HIG-16-033}
\\ \hline 
\multirow{2}{*}{ $b \bar{b}$ } & $ggF $ & $-$  & $-$  \\
& $VBF $ & $\rm -3.9^{+2.8}_{-2.9} $ \cite{ATLAS-CONF-2016-063} & $\rm -3.7^{+2.4}_{-2.5} $ \cite{CMS-PAS-HIG-16-003}  \\
\hline \hline
\end{tabular}
\caption{The Higgs signal strengths in various production and decay channels measured by
ATLAS and CMS at LHC Run 2 ($\sqrt{s}=13$ TeV).}
\label{table2}
\end{center}
\end{table}

\noindent
For the $\tau^-\tau^+$ and $W^+W^-$ channels, we used the combined results at LHC Run 1 \cite{ATLAS:2014aga,Chatrchyan:2013iaa,ATLAS-CONF-2014-061}, whereas the Higgs to diphoton signal strength at 13 TeV \cite{ATLAS:2016nke} was considered to control the variation of the previously defined observable $R_{\gamma\gamma}$.

\end{document}